\documentclass[prd,nofootinbib,onecolumn,preprint,11pt]{revtex4}

\usepackage{graphicx}
\usepackage{bm}
\usepackage{float}
\usepackage{subfigure}
\usepackage{amsmath}
\usepackage{amsfonts}
\usepackage{color}
\usepackage{slashed}
\usepackage{fancyhdr}

\newcommand{\GN}{G_{\rm N}}
\newcommand{\gb}{\bar{g}}
\newcommand{\Db}{\overline{\nabla}}

\newcommand{\sbar}{\bar{\sigma}}

\newcommand{\dd}{\text{d}}

\begin{document}

\title{Gravitational Wave Memory In dS$_{4+2n}$ and 4D Cosmology}

\author{Yi-Zen Chu}
\affiliation{
Department of Physics, University of Minnesota, 1023 University Dr., Duluth, MN 55812, USA
}
%\date{}

\begin{abstract}
\noindent We argue that massless gravitons in all even dimensional de Sitter (dS) spacetimes higher than two admit a linear memory effect arising from their propagation inside the null cone. Assume that gravitational waves (GWs) are being generated by an isolated source, and over only a finite period of time $\eta_\text{i} \leq \eta \leq \eta_\text{f}$. Outside of this time interval, suppose the shear-stress of the GW source becomes negligible relative to its energy-momentum and its mass quadrupole moments settle to static values. We then demonstrate, the transverse-traceless (TT) GW contribution to the perturbation of any dS$_{4+2n}$ written in a conformally flat form ($a^2 \eta_{\mu\nu} \dd x^\mu \dd x^\nu$) -- after the source has ceased and the primary GW train has passed -- amounts to a spacetime constant shift in the flat metric proportional to the difference between the TT parts of the source's final and initial mass quadrupole moments. As a byproduct, we present solutions to Einstein's equations linearized about de Sitter backgrounds of all dimensions greater than three. We then point out there is a similar but approximate tail induced linear GW memory effect in 4D matter dominated universes. Our work here serves to improve upon and extend the 4D cosmological results of \cite{Chu:2015yua}, which in turn preceded complementary work by Bieri, Garfinkle and Yau \cite{Bieri:2015jwa} and by Kehagias and Riotto \cite{Kehagias:2016zry}.
\end{abstract}

\maketitle

\section{Introduction and Motivation}
\label{Section_Introduction}
Consider two test masses sweeping out their respective (not necessarily geodesic) world lines in a $(d > 3)$-dimensional spatially flat Friedmann-Lema\^{i}tre-Robertson-Walker (FLRW)-like universe, namely one described by the geometry
\begin{align}
\gb_{\mu\nu} \dd x^\mu \dd x^\nu 	&\equiv a[\eta]^2 \eta_{\mu\nu} \dd x^\mu \dd x^\nu, \\
\eta_{\mu\nu} 						&\equiv \text{diag}\left[1,-1,\dots,-1\right], \qquad\qquad 
x^\mu \equiv (x^0,x^i) \equiv (\eta,\vec{x}) .
\end{align}
Suppose a gravitational wave passes by, perturbing the geometry to become
\begin{align}
g_{\mu\nu} \dd x^\mu \dd x^\nu 
\label{GeneralExpansion}
&\equiv \left( \gb_{\mu\nu} + h_{\mu\nu} \right) \dd x^\mu \dd x^\nu 						\\
\label{PerturbedFLRW}
&\equiv a[\eta]^2 \left(\eta_{\mu\nu} + \chi_{\mu\nu}[\eta,\vec{x}]\right) \dd x^\mu \dd x^\nu .
\end{align}
Denote the proper geodesic length between the two masses at a fixed time $\eta$ to be $L[\eta]$. A direct calculation, reviewed in appendix \eqref{Section_GeodesicDistance}, would then reveal that the presence of such a GW would, at leading order, yield a fractional distortion of this proper length proportional to $\chi_{ij}$.
\begin{align}
\label{FractionalDistortion}
\left(\frac{\delta L}{L}\right)[\eta,\widehat{n}]
	&= -\frac{\widehat{n}^i \widehat{n}^j}{2} \int_{0}^{1} \chi_{ij}\left[\eta,\vec{X}_0[\lambda]\right] \dd\lambda 
	+ \mathcal{O}\left[ \left(\chi_{ij}\right)^2 \right] , 
\end{align}
where $\widehat{n} \equiv (\vec{X}-\vec{X}')/|\vec{X}-\vec{X}'|$ is the unit radial vector and $\vec{X}_0[\lambda] \equiv \vec{X}' + \lambda (\vec{X}-\vec{X}')$ is a straight line in Euclidean space joining one test mass at $\vec{X}'$ to the other at $\vec{X}$. Furthermore, if after the source of the GWs has ceased and the primary GW train has gone by, the perturbation $\chi_{ij}$ does not die down to zero amplitude, but settles instead to a non-zero constant matrix $C_{ij}$, we see that the fractional distortion becomes permanent:
\begin{align}
\label{GWMemory}
\left(\frac{\delta L}{L}\right)_\text{memory} &= -\frac{\widehat{n}^i \widehat{n}^j}{2} C_{ij} + \mathcal{O}\left[ (C_{ij})^2 \right] .
\end{align}
This permanent displacement of test masses after the passage of a GW is known as the memory effect.

In this paper we shall focus primarily on background de Sitter (dS) spacetimes $\gb_{\mu\nu}$ of dimensions greater than $3$. Written in conformally flat coordinates the metric is given by
\begin{align}
\label{dS_Metric}
\gb_{\mu\nu} \dd x^\mu \dd x^\nu = a[\eta]^2 \eta_{\mu\nu} \dd x^\mu \dd x^\nu, \qquad a[\eta] \equiv -\frac{1}{H \eta}, \qquad \eta \in (-\infty,0),
\end{align}
where $H > 0$ is the Hubble parameter describing the acceleration of an expanding universe. The main thrust of this work is to argue that, the portion of GWs traveling inside the null cone of its source -- commonly known as its tail -- would give rise to a memory effect in even dimensions higher than $2$. As we shall see, this phenomenon is intimately tied to the fact that the tail part of the minimally coupled massless scalar Green's function in dS$_{4+2n}$, for $n$ zero or a positive integer, is a spacetime constant.

One may ask why a 4D physicist should be interested in physics of other dimensions. One answer is that $d$, the dimension of spacetime, may be viewed as a parameter in the equations of physics, including that of Einstein gravity. By varying $d$ we may gain insight into questions of principle, and even acquire new means to calculate 4D physics. An example is that of the tail effect itself: even though massless fields/particles propagate strictly on the light cone in 4D flat spacetimes, they no longer do so in $2$ and odd dimensions. Moreover, it is possible to understand why tails exist in odd dimensional flat spacetime by embedding it in one higher dimensional Minkowski \cite{SoodakTiersten}. This has prompted further generalizations to de Sitter spacetime \cite{Chu:2013hra}, which can be viewed as a hyperboloid situated in one higher dimensional flat spacetime. Specifically, the causal structure of the de Sitter scalar Green's function can be related to that of signals generated by an appropriately defined line source in the ambient flat spacetime. In addition, recent work \cite{Strominger:2014pwa} has drawn connections between the (better known) GW memory effect in asymptotically Minkowskian spacetime and the low frequency limit of the Ward-Takahashi identities obeyed by GW scattering amplitudes, which in turn is a consequence of the Bondi-van der Burg-Metzner-Sachs (BMS) symmetry at null infinity. Does an analogous relationship hold in cosmological spacetimes, or does it break down because the tail induced memory effect described here is timelike instead of null? These questions have first been raised in \cite{Chu:2015yua}, and subsequently in \cite{Kehagias:2016zry}; because it is one based on symmetry, it ought to be examined in all possible dimensions in order to understand the breadth of its validity.

In \S \eqref{Section_LinearizedEinstein} we work out Einstein's equations with a non-zero cosmological constant, linearized about a de Sitter background of dimensions greater or equal to four. We then express the metric perturbations as appropriate retarded Green's functions convolved against the energy-momentum-shear-stress tensor of the source(s). Following that, in \S\eqref{Section_MomentsConservation} we take a detour to define the mass and pressure quadrupole moments of an isolated GW source in a spatially flat FLRW universe, and then relate linear combinations of their time derivatives to the spatial-volume integral of the shear-stress of the same source. We then employ these results in \S \eqref{Section_MemoryEffect_dS}, and describe how the solutions laid out in \S \eqref{Section_LinearizedEinstein} lead us to a linear GW memory effect exhibited by the tensor mode, after the source has settled down. In \S \eqref{Section_MemoryEffect_4DCosmology} we re-visit some results obtained in \cite{Chu:2015yua} and comment on their implications for the linear GW memory effect in 4D cosmologies. We summarize and discuss future directions in \S \eqref{Section_SummaryFutureDirections}. In appendix \S\eqref{Section_GeodesicDistance} we review the calculation of spatial geodesic distances between a pair of test masses in a perturbed spatially flat FLRW-like universe. In appendix \S\eqref{Section_PDESolution} we delineate the solutions of the partial differential equations arising from General Relativity linearized about a background de Sitter spacetime. Despite being an appendix, this section is the technical heart of the paper. Finally, in appendix \eqref{Section_Bardeen} we identify the gauge-invariant metric perturbation variables in a $(d \geq 4)$-dimensional background spatially flat FLRW geometry.

\section{General Relativity With $\Lambda$, Linearized About (de Sitter)$_{d \geq 4}$}
\label{Section_LinearizedEinstein}
Einstein's equations for the metric $g_{\mu\nu}$, sourced by the energy-momentum-shear-stress tensor $T_{\mu\nu}$ of some matter source, is
\begin{align}
\label{Einstein}
G_{\mu\nu}[g] - \Lambda g_{\mu\nu} 
= R_{\mu\nu}[g] - \frac{1}{2} g_{\mu\nu} \mathcal{R}[g] - \Lambda g_{\mu\nu} 
= 8 \pi \GN T_{\mu\nu} .
\end{align}
$\GN$ is Newton's gravitational constant in $d$ dimensions. We are including a positive cosmological constant $\Lambda > 0$ because mounting astrophysical evidence points to its existence in the 4D universe we reside in. Moreover, for this paper, we will assume that $T_{\mu\nu}$ describes an isolated, compact astrophysical system which does not distort the overall geometry too much. When we neglect its influence (i.e., set $T_{\mu\nu}=0$) we recover the pure de Sitter spacetime in eq. \eqref{dS_Metric}, which in turn solves
\begin{align}
\label{dS_EOM}
G_{\mu\nu}[\gb] - \Lambda \gb_{\mu\nu} = 0 .
\end{align}
Our approach to solving eq. \eqref{Einstein} is then a perturbative one, through eq. \eqref{PerturbedFLRW}, by expanding the geometry about $\gb_{\mu\nu}$. 

{\bf General Perturbation Theory (PT)} \qquad To this end, it is convenient to work with the barred graviton, namely
\begin{align}
\overline{h}_{\mu\nu} \equiv h_{\mu\nu} - \frac{\gb_{\mu\nu}}{2} \gb^{\alpha\beta} h_{\alpha\beta} .
\end{align}
When $\gb_{\mu\nu} = \eta_{\mu\nu}$ in 4D, this $\bar{h}_{\mu\nu}$ is more commonly known as the ``trace-reversed" graviton. Equivalently, from eq. \eqref{PerturbedFLRW}, we may define
\begin{align}
\overline{\chi}_{\mu\nu} \equiv \chi_{\mu\nu} - \frac{\eta_{\mu\nu}}{2} \eta^{\alpha\beta} \chi_{\alpha\beta} .
\end{align}
Without specializing to the perturbed FLRW form in eq. \eqref{PerturbedFLRW}, any expansion about some general background metric $\gb_{\mu\nu}$ given in eq. \eqref{GeneralExpansion} would yield the following left hand side of Einstein's eq. \eqref{Einstein}:
\begin{align}
\label{Einstein_LHSExpansion}
&G_{\mu\nu}[g] - \Lambda g_{\mu\nu} = G_{\mu\nu}[\gb] - \Lambda \gb_{\mu\nu} \\
&-\frac{1}{2} \left( \overline{\Box} \bar{h}_{\mu\nu} - \Db_{\{\mu} \Db^\alpha \bar{h}_{\nu\} \alpha} + \gb_{\mu\nu} \Db_\alpha \Db_\beta \bar{h}^{\alpha\beta} 
-\gb_{\mu\nu} \bar{h}^{\alpha\beta} R_{\alpha\beta}[\gb] - \bar{h}_{\{ \mu}^{\phantom{\{\mu} \sigma} R_{\nu\} \sigma}[\gb] + \bar{h}_{\mu\nu} \mathcal{R}[\gb] + 2 \bar{h}^{\alpha\beta} R_{\mu\alpha\nu\beta}[\gb] \right) \nonumber\\
&\qquad\qquad\qquad\qquad
- \Lambda \left( \bar{h}_{\mu\nu} - \frac{\gb_{\mu\nu}}{d-2} \gb^{\alpha\beta} \bar{h}_{\alpha\beta} \right) + \mathcal{O}[\bar{h}^2] ,
\qquad\qquad \overline{\Box} \equiv \Db_\sigma \Db^\sigma . \nonumber
\end{align}
In eq. \eqref{Einstein_LHSExpansion}, the geometric tensors -- Einstein $G_{\mu\nu}[\gb]$, Ricci $R_{\mu\nu}[\gb]$, Ricci scalar $\mathcal{R}[\gb]$ and Riemann $R_{\mu\alpha\nu\beta}[\gb]$ -- and the covariant derivative $\Db$ are built solely out of the background $\gb_{\mu\nu}$; moreover, all indices are moved with it. The symmetrization symbol $\{ \dots \}$ is defined through the relation $T_{\{ \alpha \beta \}} \equiv T_{\alpha\beta} + T_{\beta\alpha}$.

{\bf PT about dS$_{d \geq 4}$} \qquad Now, if we do specialize to $\gb_{\mu\nu}$ being the $d$ dimensional de Sitter metric in eq. \eqref{dS_Metric}, which satisfies eq. \eqref{dS_EOM}, the first line on the right hand side of eq. \eqref{Einstein_LHSExpansion} vanishes. We then use the maximally symmetric form that the de Sitter geometric tensors take, namely
\begin{align}
R_{\mu\nu \alpha\beta}[\gb] = - \frac{2}{(d-1)(d-2)} \Lambda \left( \gb_{\mu\alpha} \gb_{\beta\nu} - \gb_{\mu\beta} \gb_{\alpha\nu} \right), \\
R_{\alpha\beta}[\gb] = - \frac{2}{d-2} \Lambda \gb_{\alpha\beta}, \qquad
\mathcal{R}[\gb] = - \frac{2 d}{d-2} \Lambda ,
\end{align}
followed by the relationship between the Hubble parameter $H$ in eq. \eqref{dS_Metric} and the cosmological constant $\Lambda$ in eq. \eqref{Einstein},
\begin{align}
\Lambda = \frac{H^2}{2} (d-1)(d-2) .
\end{align}
Evaluated on a de Sitter background geometry, eq. \eqref{Einstein_LHSExpansion} then bring us to
\begin{align}
\label{Einstein_LHSExpansion_dS}
G_{\mu\nu}[g] - \Lambda g_{\mu\nu}
&= \frac{1}{2} \left(
- \overline{\Box} \bar{h}_{\mu\nu} + \Db_{\{\mu} \Db^\alpha \bar{h}_{\nu\}\alpha} - \gb_{\mu\nu} \Db_\alpha \Db_\beta \overline{h}^{\alpha\beta} \right) + H^2 \left( \gb_{\mu\nu} \gb^{\alpha\beta} \overline{h}_{\alpha\beta} - \overline{h}_{\mu\nu} \right) + \mathcal{O}[\bar{h}^2] . 
\end{align}
To solve Einstein's eq. \eqref{Einstein} perturbatively, we need to choose a gauge for the barred graviton field $\overline{h}_{\alpha\beta}$, so that its wave operator can be inverted and its solution written as a convolution of appropriate Green's functions against the sources $8 \pi \GN T_{\alpha\beta}$.

{\it Gauge fixing} \qquad We now require $\bar{h}_{\alpha\beta}$ to satisfy the gauge condition
\begin{align}
\Db^\sigma \bar{h}_{\sigma\nu} = \Db^\sigma \ln\left[ a^2 \right] \bar{h}_{\sigma\nu} .
\end{align}
Equivalently, in terms of $\bar{\chi}_{\alpha\beta}$ (recall equations \eqref{GeneralExpansion} and \eqref{PerturbedFLRW}),
\begin{align}
\label{dS_GaugeFixing}
\partial^\mu \overline{\chi}_{\mu\nu} = \frac{1}{\eta} \left( (d-2) \overline{\chi}_{0\nu} - \delta_\nu^0 \eta^{\alpha\beta} \overline{\chi}_{\alpha\beta} \right) ,
\end{align}
where all indices in eq. \eqref{dS_GaugeFixing} and in what follows are moved with the flat metric $\eta_{\mu\nu}$. This gauge in eq. \eqref{dS_GaugeFixing} and the ensuing equations below, are really a generalization of the 4D ones in \cite{deVega:1998ia}, \cite{Ashtekar:2015lxa} and \cite{Date:2015kma}. At this point, eq. \eqref{Einstein_LHSExpansion_dS} becomes
\begin{align}
G_{\mu\nu}[g] - \Lambda g_{\mu\nu}
&= -\frac{1}{2} \left\{ 
a^2 \overline{\Box}^\text{(S)} \overline{\chi}_{\mu\nu} 
- \frac{1}{\eta^2} \left( 2 \delta_\mu^0 \delta_\nu^0 \eta^{\sigma\rho} \overline{\chi}_{\sigma\rho} - (d-2) \overline{\chi}_{0 \{ \mu} \delta_{\nu\}}^0 \right)
\right\}
+ \mathcal{O}[\overline{\chi}^2] , 
\end{align}
where the $\overline{\Box}^\text{(S)} = \Db_\alpha \Db^\alpha$ is the scalar one with respect to the background de Sitter metric, namely
\begin{align}
\overline{\Box}^\text{(S)} \overline{\chi}_{\mu\nu} 
= \frac{\partial_\alpha \left( \sqrt{|\gb|} \gb^{\alpha\beta} \partial_\beta \overline{\chi}_{\mu\nu} \right)}{\sqrt{|\gb|}}
= a^{-2} \left( \partial^2 \overline{\chi}_{\mu\nu} - \frac{d-2}{\eta} \partial_0 \overline{\chi}_{\mu\nu} \right) .
\end{align}
(The $\partial^2 \equiv \eta^{\mu\nu} \partial_\mu \partial_\nu$ is the wave operator in flat spacetime.) The linearized version of eq. \eqref{Einstein} is thus
\begin{align}
\label{EinsteinLinearized}
a^2 \overline{\Box}^\text{(S)} \overline{\chi}_{\mu\nu} 
- \frac{1}{\eta^2} \left( 2 \delta_\mu^0 \delta_\nu^0 \overline{\chi} - (d-2) \overline{\chi}_{0 \{ \mu} \delta_{\nu\}}^0 \right)
= -16\pi\GN T_{\mu\nu}, \qquad\qquad
\overline{\chi} \equiv \eta^{\sigma\rho} \overline{\chi}_{\sigma\rho} .
\end{align}
The solution of eq. \eqref{EinsteinLinearized} is the primary technical focus of this work.

{\it Pseudo-trace mode} \qquad By adding $(3-d)$ times of the $00$ component of the linearized Einstein's equation \eqref{EinsteinLinearized} to its spatial-trace, we are lead to
\begin{align}
\label{dS_PseudoTrace_IofII}
\partial^2 \widetilde{\overline{\chi}} - \frac{d-2}{\eta} \partial_0 \widetilde{\overline{\chi}}
- \frac{2}{\eta^2} (3-d) \widetilde{\overline{\chi}} &= -16\pi \GN \widetilde{T}, 
\end{align}
where
\begin{align}
\widetilde{\overline{\chi}} &\equiv (3-d) \overline{\chi}_{00} - \delta^{ij} \overline{\chi}_{ij} , \\
\widetilde{T} 				&\equiv (3-d) T_{00} - \delta^{ij} T_{ij} .
\end{align}
An equivalent form of eq. \eqref{dS_PseudoTrace_IofII} is
\begin{align}
\label{dS_PseudoTrace_IIofII}
\left( \partial^2 - \frac{(d-6)(d-4)}{4 \eta^2} \right) \left( a^{\frac{d-2}{2}} \widetilde{\overline{\chi}} \right)
= -16\pi \GN a^{\frac{d-2}{2}} \widetilde{T} .
\end{align}
Here and below -- equations \eqref{dS_PseudoTrace_IIofII}, \eqref{dS_Vector_IIofII} and \eqref{dS_Tensor_IIofII} -- because we are faced with partial differential equations (PDEs) of the same form, namely $(\partial^2+U[\eta])(a^{\frac{d-2}{2}}\psi) = -16\pi \GN a^{\frac{d-2}{2}} S$, we devote appendix \eqref{Section_PDESolution} to solving the relevant Green's functions. In this section we will merely quote the final results. The pseudo-trace retarded solutions are
\begin{align}
\label{dS_PseudoTrace_Solution}
\widetilde{\overline{\chi}}[\eta,\vec{x}]
&= -16\pi \GN H^{d-2} \int_{-\infty}^{\eta} \dd\eta' \int_{\mathbb{R}^{d-1}} \dd^{d-1}\vec{x}' a[\eta']^{d-2} 
\mathcal{G}^{\text{(Tr)}}\left[ s \right] \widetilde{T}[\eta',\vec{x}'] , \\
s &\equiv \frac{\sbar}{\eta\eta'}, \qquad \qquad
		\sbar = \frac{1}{2} \eta_{\mu\nu} (x-x')^\mu (x-x')^\nu \equiv \frac{1}{2}(x-x')^2 ,
\end{align}
with $\sbar$ being Synge's world function in Minkowski spacetime; and
\begin{align}
\mathcal{G}^{\text{(Tr)}}_{\text{even $d \geq 4$}}\left[ s \right]
&= \frac{1}{(2\pi)^{\frac{d-2}{2}}} \left( \frac{\partial}{\partial s} \right)^{\frac{d-2}{2}} \left( \frac{\Theta[s]}{2} P_{\frac{d-6}{2}}[1+s] \right) , \\
\mathcal{G}^{\text{(Tr)}}_{\text{odd $d \geq 5$}}\left[ s \right]
&= \frac{1}{(2\pi)^{\frac{d-3}{2}}} \left( \frac{\partial}{\partial s} \right)^{\frac{d-3}{2}} \left( \frac{\Theta[s]}{4\pi} \frac{\left(s+\sqrt{s(s+2)}+1\right)^{d-5}+1}{\sqrt{s(s+2)} \left(s+\sqrt{s (s+2)}+1\right)^{\frac{d-5}{2}}} \right) .
\end{align}
{\it Vector mode} \qquad The $0i$ components of eq. \eqref{EinsteinLinearized} reads
\begin{align}
\label{dS_Vector_IofII}
\partial^2 \overline{\chi}_{0i} - \frac{d-2}{\eta} \partial_0 \overline{\chi}_{0i} + \frac{d-2}{\eta^2} \overline{\chi}_{0i} &= -16\pi \GN T_{0i} , 
\end{align}
or
\begin{align}
\label{dS_Vector_IIofII}
\left( \partial^2 - \frac{(d-4)(d-2)}{4 \eta^2} \right) \left( a^{\frac{d-2}{2}} \overline{\chi}_{0i} \right)
= -16\pi \GN a^{\frac{d-2}{2}} T_{0i} .
\end{align}
The retarded solutions are
\begin{align}
\label{dS_Vector_Solution}
\overline{\chi}_{0i}[\eta,\vec{x}]
&= -16\pi \GN H^{d-2} \int_{-\infty}^{\eta} \dd\eta' \int_{\mathbb{R}^{d-1}} \dd^{d-1}\vec{x}' a[\eta']^{d-2} 
\mathcal{G}^{\text{(V)}}\left[ s \right] T_{0i}[\eta',\vec{x}'] ,
\end{align}
where 
\begin{align}
\mathcal{G}^{\text{(V)}}_{\text{even $d \geq 4$}}\left[ s \right]
&= \frac{1}{(2\pi)^{\frac{d-2}{2}}} \left( \frac{\partial}{\partial s} \right)^{\frac{d-2}{2}} \left( \frac{\Theta[s]}{2} P_{\frac{d-4}{2}}[1+s] \right) , \\
\mathcal{G}^{\text{(V)}}_{\text{odd $d \geq 5$}}\left[ s \right]
&= \frac{1}{(2\pi)^{\frac{d-3}{2}}} \left( \frac{\partial}{\partial s} \right)^{\frac{d-3}{2}} \left( \frac{\Theta[s]}{4\pi} \frac{\left(s+\sqrt{s(s+2)}+1\right)^{d-3}+1}{\sqrt{s(s+2)} \left(s+\sqrt{s (s+2)}+1\right)^{\frac{d-3}{2}}} \right) .
\end{align}
{\it Tensor mode} \qquad The equations for the $ij$ components of eq. \eqref{EinsteinLinearized} turn out to be that of the minimally coupled massless scalar in de Sitter spacetime,
\begin{align}
\label{dS_Tensor_IofII}
\overline{\Box}^\text{(S)} \overline{\chi}_{ij} = -\frac{16\pi \GN}{a^2} T_{ij} .
\end{align}
This translates to
\begin{align}
\label{dS_Tensor_IIofII}
\left( \partial^2 - \frac{d(d-2)}{4 \eta^2} \right) \left( a^{\frac{d-2}{2}} \overline{\chi}_{ij} \right)
= -16\pi \GN a^{\frac{d-2}{2}} T_{ij} .
\end{align}
The retarded solutions are
\begin{align}
\label{dS_Tensor_Solution}
\overline{\chi}_{ij}[\eta,\vec{x}]
&= -16\pi \GN H^{d-2} \int_{-\infty}^{\eta} \dd\eta' \int_{\mathbb{R}^{d-1}} \dd^{d-1}\vec{x}' a[\eta']^{d-2} 
\mathcal{G}^{\text{(T)}}\left[ s \right] T_{ij}[\eta',\vec{x}'] ,
\end{align}
where 
\begin{align}
\mathcal{G}^{\text{(T)}}_{\text{even $d \geq 4$}}\left[ s \right]
&= \frac{1}{(2\pi)^{\frac{d-2}{2}}} \left( \frac{\partial}{\partial s} \right)^{\frac{d-2}{2}} \left( \frac{\Theta[s]}{2} P_{\frac{d-2}{2}}[1+s] \right) , \\
\mathcal{G}^{\text{(T)}}_{\text{odd $d \geq 5$}}\left[ s \right] 
&= \frac{1}{(2\pi)^{\frac{d-3}{2}}} \left( \frac{\partial}{\partial s} \right)^{\frac{d-3}{2}} \left( \frac{\Theta[s]}{4\pi} \frac{\left(s+\sqrt{s(s+2)}+1\right)^{d-1}+1}{\sqrt{s(s+2)} \left(s+\sqrt{s (s+2)}+1\right)^{\frac{d-1}{2}}} \right) .
\end{align}
We highlight here that, these solutions in eq. \eqref{dS_Tensor_Solution} amount to the convolution of $-16 \pi \GN T_{ij}/a^2$ against the minimally coupled massless scalar Green's function in de Sitter spacetime. As we shall witness, the latter's spacetime constant tail in even ($d \geq 4$)-dimensional spacetime is responsible for contributing to the linear GW memory effect.

\section{Mass and pressure quadrupole moments; Conservation laws in a spatially flat FLRW spacetime}
\label{Section_MomentsConservation}

Before examining this tail induced linear GW memory effect arising from the even $d \geq 4$ solutions in eq. \eqref{dS_Tensor_Solution}, however, we need to first -- following \cite{Ashtekar:2015lxa} and \cite{Date:2015kma} -- relate spatial-volume integrals of the shear-stress $T_{ij}$ of the isolated matter source to its mass $Q_{ij}$ and pressure $P_{ij}$ quadrupole moments.

{\bf Quadrupole moments} \qquad Throughout this section, unless otherwise indicated, we will suppose our background metric is a spatially flat FLRW universe, i.e., not necessarily de Sitter:
\begin{align}
\label{SpatiallyFlatFLRW}
\gb_{\mu\nu} \dd x^\mu \dd x^\nu = a[\eta]^2 \eta_{\mu\nu} .
\end{align}
We then note that the $d$-beins $\{ \varepsilon^{\mu}_{\phantom{\mu}\widehat{\alpha}} \}$ of such a spacetime, whose defining property is
\begin{align}
\varepsilon^{\mu}_{\phantom{\mu}\widehat{\alpha}} \varepsilon^{\nu}_{\phantom{\nu}\widehat{\beta}} \gb_{\mu\nu} = \eta_{\alpha\beta} ,
\end{align}
are given by
\begin{align}
\varepsilon^{\mu}_{\phantom{\mu}\widehat{\nu}} = a^{-1} \delta^\mu_{\phantom{\mu}\nu} .
\end{align}
The (upper) $\mu$ index of $\varepsilon^{\mu}_{\phantom{\mu}\widehat{\nu}}$ transforms as a coordinate vector; while its (lower) $\widehat{\nu}$ index transforms as a local Lorentz $1$-form. Therefore we can form from $T_{\mu\nu}$ the $d \times d$ matrix of coordinate scalar quantities
\begin{align}
\label{OrthonormalFrame}
T_{\widehat{\alpha}\widehat{\beta}} 
\equiv \varepsilon^{\mu}_{\phantom{\mu}\widehat{\alpha}} \varepsilon^{\nu}_{\phantom{\nu}\widehat{\beta}} T_{\mu\nu}
= a^{-2} T_{\alpha\beta} .
\end{align}
The $T_{\widehat{0}\widehat{0}}$ can now be interpreted as the mass-energy density measured by a local observer; $T_{\widehat{0}\widehat{i}}$ as its $(d-1)$-momentum density; and $T_{\widehat{i}\widehat{j}}$ as its shear-stress/pressure density. Moreover, from eq. \eqref{SpatiallyFlatFLRW}, since the induced metric on a constant$-\eta$ hypersurface is $\dd\vec{\ell}^2 = -a^2 \delta_{ij} \dd x^i \dd x^j$, we may recognize the proper spatial volume on the said hypersurface to be
\begin{align}
\dd^{d-1}(\text{proper vol.})[\eta] = \dd^{d-1}\vec{x} a^{d-1}[\eta] .
\end{align}
The physical mass and pressure quadrupole moments are now defined as
{\allowdisplaybreaks\begin{align}
\label{MassQuadrupole}
Q^{ij}[\eta] = Q_{ij}[\eta] 
&\equiv \int_{\mathbb{R}^{d-1}} \dd^{d-1} \vec{x} a^{d-1} (a x^i) (a x^j) T_{\widehat{0}\widehat{0}}[\eta,\vec{x}] ,  \\
&= a^{d-1} \int_{\mathbb{R}^{d-1}} \dd^{d-1} \vec{x} x^i x^j T_{00}[\eta,\vec{x}], \nonumber\\
\label{PressureQuadrupole}
P^{ij}[\eta] = P_{ij}[\eta] 
&\equiv \int_{\mathbb{R}^{d-1}} \dd^{d-1} \vec{x} a^{d-1} (a x^i) (a x^j) \delta^{ab} T_{\widehat{a}\widehat{b}}[\eta,\vec{x}] , \\
&= a^{d-1} \int_{\mathbb{R}^{d-1}} \dd^{d-1} \vec{x} x^i x^j \delta^{ab} T_{ab}[\eta,\vec{x}] , \nonumber
\end{align}}
where we associate one scale factor to each of the $x^i$ and $x^j$ to form a physical vector.

{\bf Conservation} \qquad At linear order, the energy-momentum-shear-stress tensor $T_{\mu\nu}$ of an isolated astrophysical system in a spatially flat FLRW background geometry is conserved
\begin{align}
\Db^\mu T_{\mu\nu} = 0 ,
\end{align}
with $\Db$ being the covariant derivative with respect to the $\gb_{\mu\nu}$. (This statement, of course, neglects backreaction.) A direct calculation reveals
\begin{align}
\label{Conservation_I_0}
\partial_0 \left( a^{d-2} T_{00} \right) 
	&= \delta^{lj} \partial_l \left( a^{d-2} T_{j0} \right) + a^{d-2} \frac{\dot{a}}{a} \eta^{\alpha\beta} T_{\alpha\beta} , \\
\label{Conservation_I_i}
\partial_0 \left( a^{d-2} T_{0i} \right) 
	&= \delta^{lj} \partial_l \left( a^{d-2} T_{ji} \right) .
\end{align}
Differentiating both sides of eq. \eqref{Conservation_I_0} once with respect to time and employing eq. \eqref{Conservation_I_i} on the resulting right hand side leads us to
\begin{align}
\label{Conservation_II}
\partial_0 \left\{ \partial_0 \left( a^{d-2} T_{00} \right) - \frac{\dot{a}}{a} a^{d-2} (T_{00} - \delta^{lj} T_{lj}) \right\}
= \delta^{lk} \delta^{ij} \partial_l \partial_i \left( a^{d-2} T_{kj} \right) .
\end{align}
We may use eq. \eqref{Conservation_II} in the following way. Via integration-by-parts and the assumption that the matter distribution is localized in space (so that surface terms are zero) -- one may readily see that
\begin{align}
\frac{1}{2} \int_{\mathbb{R}^{d-1}} \dd^{d-1} \vec{x} x^a x^b \delta^{lk} \delta^{ij} \partial_l \partial_i \left( a^{d-2} T_{kj} \right)
= \int_{\mathbb{R}^{d-1}} \dd^{d-1} \vec{x} a^{d-2} T_{ab} .
\end{align}
Applying eq. \eqref{Conservation_II} to the left hand side allow us to arrive at
\begin{align}
\label{Conservation_III}
\int_{\mathbb{R}^{d-1}} \dd^{d-1} \vec{x} a[\eta]^{d-2} T_{ab}[\eta,\vec{x}]
&= \frac{1}{2} \partial_0 \left\{
\frac{\partial_0 Q_{ab}[\eta]}{a[\eta]} - \frac{\dot{a}[\eta]}{a[\eta]^2} \left(2 Q_{ab}[\eta] - P_{ab}[\eta]\right) \right\} ,
\end{align}
where we have inserted the mass and pressure quadrupole definitions from equations \eqref{MassQuadrupole} and \eqref{PressureQuadrupole}. We reiterate that eq. \eqref{Conservation_III} holds in any $d$-dimensional spatially flat FLRW geometry. When we specialize to de Sitter spacetime, where $a[\eta] = -1/(H\eta)$,
\begin{align}
\label{Conservation_dS}
\int_{\mathbb{R}^{d-1}} \dd^{d-1} \vec{x} a^{d-2} T_{ab}
&= \frac{1}{2} \partial_0 \left\{ \frac{\partial_0 Q_{ab}}{a} - H \left( 2 Q_{ab} - P_{ab} \right) \right\} , \qquad\qquad \text{(de Sitter)} .
\end{align}

\section{Tail Induced Linear GW Memory Effect in (de Sitter)$_{4+2n}$}
\label{Section_MemoryEffect_dS}
We now turn our attention to the tail part of the metric perturbations in even dimensional ($d \geq 4$) de Sitter spacetime, as encoded in equations \eqref{dS_PseudoTrace_Solution}, \eqref{dS_Vector_Solution} and \eqref{dS_Tensor_Solution}. Despite experiencing a non-trivial potential in higher dimensions (cf. equations \eqref{dS_PseudoTrace_IIofII} and \eqref{dS_Vector_IIofII}), the pseudo-trace $\widetilde{\overline{\chi}}$ and vector $\overline{\chi}_{0i}$ exhibit no tails. The physical reason is unclear; however, the mathematical reason is that $P_{(d-6)/2}[1+s]$ and $P_{(d-4)/2}[1+s]$ are respectively polynomials in $s$ of degree $((d-2)/2)-2$ and $((d-2)/2)-1$. Therefore,
\begin{align}
\label{dS_PseudoTrace_Even_NoTails}
\mathcal{G}^{\text{(Tr$\vert$tail)}}_{\text{even $d \geq 4$}}\left[ s \right]
&= \frac{\Theta[s]}{2 (2\pi)^{\frac{d-2}{2}}} \left( \frac{\partial}{\partial s} \right)^{\frac{d-2}{2}} P_{\frac{d-6}{2}}[1+s] = 0 , \\
\label{dS_Vector_Even_NoTails}
\mathcal{G}^{\text{(V$\vert$tail)}}_{\text{even $d \geq 4$}}\left[ s \right]
&= \frac{\Theta[s]}{2 (2\pi)^{\frac{d-2}{2}}} \left( \frac{\partial}{\partial s} \right)^{\frac{d-2}{2}} P_{\frac{d-4}{2}}[1+s] = 0 .
\end{align}
(We have checked the second line against the 6D light cone boundary condition in eq. \eqref{G_6D_LCBC}, i.e., it is zero for $p=(d-4)(d-2)/4$.) In the gauge of eq. \eqref{dS_GaugeFixing}, it is thus only the tensor mode $\overline{\chi}_{ij}$ that travels inside the light cone of its source. Moreover, the tail of its Green's function $\mathcal{G}^{\text{(T)}}$ in eq. \eqref{dS_Tensor_Solution} is a constant because one is differentiating a $(d-2)/2$ degree polynomial $(d-2)/2$ times,
\begin{align}
\mathcal{G}^{\text{(T$\vert$tail)}}_{\text{even $d \geq 4$}}\left[ s \right]
&= \frac{\Theta[s]}{2(2\pi)^{\frac{d-2}{2}}} \left(\frac{\partial}{\partial s}\right)^{\frac{d-2}{2}} P_{\frac{d-2}{2}}[1+s] 
= \frac{\Theta[s]}{2(2\pi)^{\frac{d-2}{2}}} \frac{(d-2)!}{2^{\frac{d-2}{2}} \left(\frac{d-2}{2}\right)!} .
\end{align}
(This result follows from Rodrigues' formula for the Legendre polynomials.) This is, of course, equivalent to the fact that the tail of the de Sitter minimally coupled massless scalar Green's function is a constant in all even dimensions higher than $2$ \cite{Chu:2013hra}. In even $d \geq 4$ dimensions, the tail part of the tensor mode solution in eq. \eqref{dS_Tensor_Solution} is therefore
\begin{align}
\label{dS_Even_Tails}
\overline{\chi}_{ij}^{\text{(tail)}}[\eta,\vec{x}]
&= -16\pi \GN \frac{H^{d-2}}{(2\pi)^{\frac{d-2}{2}}} \frac{(d-2)!}{2^{\frac{d}{2}} \left(\frac{d-2}{2}\right)!} 
\int_{\mathbb{R}^{d-1}} \dd^{d-1}\vec{x}' \int_{-\infty}^{\eta_r} \dd\eta' a[\eta']^{d-2} T_{ij}[\eta',\vec{x}'] , 
\end{align}
where we have defined the retarded time $\eta_r \equiv \eta-|\vec{x}-\vec{x}'|-0^+$. For technical convenience, throughout the rest of this paper, we will assume that the coordinate system has been chosen such that $\vec{x}=\vec{x}'=0$ is located within the source. If we now invoke eq. \eqref{Conservation_dS} derived in the previous section, the linear GW tail can be expressed in terms of the mass and pressure quadrupole moments of the source, at least in the far zone ($|\vec{x}| \gg |\vec{x}'|$) where $|\vec{x}-\vec{x}'| \approx |\vec{x}|$ and hence $\eta_r \approx \eta-|\vec{x}|$:
\begin{align}
\label{dS_Even_Tails_QuadrupoleMoments}
\overline{\chi}_{ij}^{\text{(tail)}}[\eta,\vec{x}]
&\approx -8 \pi \GN \frac{H^{d-2}}{(2\pi)^{\frac{d-2}{2}}} \frac{(d-2)!}{2^{\frac{d}{2}} \left(\frac{d-2}{2}\right)!} 
\left[ \frac{\partial_0 Q_{ab}[\eta']}{a[\eta']} - H \left( 2 Q_{ab}[\eta'] - P_{ab}[\eta'] \right) \right]_{\eta'=-\infty}^{\eta'=\eta_r} .
\end{align}
{\bf Negligible shear-stress, settling of quadrupole moments} \qquad Let us suppose that the isolated source is active only over a finite interval of time, $\eta_\text{i} \leq \eta \leq \eta_\text{f}$. This means we will assume that outside this interval, its shear-stress is negligible relative to its energy-momentum, so that we may set the former to zero, namely
\begin{align}
\label{ShearStressZero}
T_{ij}[\eta < \eta_{\text{i}}] = T_{ij}[\eta > \eta_{\text{f}}] = 0 .
\end{align}
This often coincides with the non-relativistic limit, achieved when a system has settled down. Equations \eqref{ShearStressZero} and \eqref{PressureQuadrupole} immediately imply the pressure quadrupole moment is zero outside the interval $\eta_\text{i} \leq \eta \leq \eta_\text{f}$:
\begin{align}
\label{PressureQuadrupoleZero}
P_{ij}[\eta < \eta_{\text{i}}] = P_{ij}[\eta > \eta_{\text{f}}] = 0 .
\end{align}
We will further assume that the mass quadrupole moment is not static only during this active interval:
\begin{align}
\label{MassQuadrupoleDotZero}
\dot{Q}_{ij}[\eta < \eta_{\text{i}}] = \dot{Q}_{ij}[\eta > \eta_{\text{f}}] = 0 .
\end{align}
Altogether, equations \eqref{ShearStressZero}, \eqref{PressureQuadrupoleZero} and \eqref{MassQuadrupoleDotZero} applied to eq. \eqref{dS_Even_Tails_QuadrupoleMoments} lead us to a tail induced linear GW memory effect. Specifically for late times ($\eta_r > \eta_\text{f}$),\footnote{The retarded time $\eta_r \equiv \eta - |\vec{x}-\vec{x}'|$ depends on $\vec{x}'$ and therefore takes part in the spatial-volume integral of eq. \eqref{dS_Even_Tails}. This is why the far zone limit, replacing $|\vec{x}-\vec{x}'| \to |\vec{x}|$, was required when transitioning to eq. \eqref{dS_Even_Tails_QuadrupoleMoments}. However, once the source has settled down, equations \eqref{ShearStressZero} and \eqref{PressureQuadrupoleZero} allow us to deduce that for $\eta_r > \eta_\text{f}$, eq. \eqref{dS_Even_Tails_QuadrupoleMoments} now holds everywhere within the future null cone of the GW source's world tube.} the assumption in eq. \eqref{ShearStressZero} allows us to perform the $\eta'-$integral in eq. \eqref{dS_Even_Tails} only over $\eta_\text{i} \leq \eta' \leq \eta_\text{f}$ and sets to zero $P_{ab}$ in eq. \eqref{Conservation_dS}; while eq. \eqref{MassQuadrupoleDotZero} puts $\partial_0 Q_{ab}=0$ in eq. \eqref{Conservation_dS}:
\begin{align}
\overline{\chi}_{ij}^{\text{(tail)}}[\eta - r > \eta_{\text{f}},\vec{x}]
&= -16\pi \GN \frac{H^{d-2}}{(2\pi)^{\frac{d-2}{2}}} \frac{(d-2)!}{2^{\frac{d}{2}} \left(\frac{d-2}{2}\right)!} 
\int_{\mathbb{R}^{d-1}} \dd^{d-1}\vec{x}' \int_{\eta_\text{i}}^{\eta_\text{f}} \dd\eta' a[\eta']^{d-2} T_{ij}[\eta',\vec{x}'] \nonumber\\
\label{dS_Even_Tails_Final}
&= 16\pi \GN \frac{H^{d-1}}{(2\pi)^{\frac{d-2}{2}}} \frac{(d-2)!}{2^{\frac{d}{2}} \left(\frac{d-2}{2}\right)!} 
\left( Q_{ij}[\eta_{\text{f}}] - Q_{ij}[\eta_{\text{i}}] \right) .
\end{align}
Therefore, after the GW source has settled down, we see that in all background dS$_{4+2n}$, the $C_{ij}$ occurring in the fractional distortion of eq. \eqref{FractionalDistortion} receives contributions solely from the $\overline{\chi}_{ij}^{\text{(tail)}}$ in eq. \eqref{dS_Even_Tails_Final}.

{\it Gravitational radiation} \qquad It is important to record here that the transverse-traceless portion of $\overline{\chi}_{ij}$ is gauge-invariant, namely $D_{ij} \equiv \overline{\chi}_{ij}^{\text{TT}}$ does not change its form under infinitesimal coordinate transformations, and is in fact what is usually meant by gravitational radiation -- like $\overline{\chi}_{ij}$ itself, it obeys the de Sitter minimally coupled massless scalar wave equation. (For the reader's reference, in appendix \eqref{Section_Bardeen} we identify the SO$_{d-1}$ scalar, vector and tensor gauge-invariant metric variables in a $d$-dimensional spatially flat FLRW geometry.) Assuming equations \eqref{ShearStressZero} and \eqref{MassQuadrupoleDotZero} hold, the assertion that the GW tail in dS$_{4+2n}$ remains a non-zero spacetime constant, after its source has ceased, is therefore a coordinate-invariant and physical one.
\begin{align}
\label{dS_Memory}
D_{ij}^{\text{(tail)}}[\eta - r > \eta_{\text{f}},\vec{x}] 
&\equiv \overline{\chi}_{ij}^{\text{(TT$\vert$tail)}}[\eta - r > \eta_{\text{f}},\vec{x}] \nonumber\\
&= 16\pi \GN \frac{H^{d-1}}{(2\pi)^{\frac{d-2}{2}}} \frac{(d-2)!}{2^{\frac{d}{2}} \left(\frac{d-2}{2}\right)!} 
\left( Q_{ij}^{\text{TT}}[\eta_{\text{f}}] - Q_{ij}^{\text{TT}}[\eta_{\text{i}}] \right) .
\end{align}
Because the GW tail here is a constant in spacetime, notice this linear GW memory effect does not decay with distance from the source, unlike the expected $1/(\text{spatial distance})^{d-3}$ fall-off in even ($d \geq 4$)-dimensional flat spacetime. Moreover, in even dimensional Minkowski spacetimes higher than $4$, linear massless gravitons continue to propagate strictly on the null cone, like their 4D cousins. For these reasons, the dS$_{4+2n}$ tail induced linear GW memory effect captured in eq. \eqref{dS_Memory} really has no counterpart in the asymptotically flat case.

{\bf Remark I} \qquad Before shifting our attention to 4D cosmology, we mention here that \cite{Ashtekar:2015lxa} advocates exploiting the Killing vector in de Sitter spacetime
\begin{align}
\label{KillingVector}
T^\mu \partial_\mu \equiv -H x^\mu \partial_\mu 
\end{align}
to define what it means for a GW source to settle down. (This $T^\mu \partial_\mu$ is timelike in the region $x^2 \equiv \eta_{\alpha\beta} x^\alpha x^\beta > 0$.) Specifically, they required that the Lie derivative of the energy-momentum-shear-stress tensor $T_{\alpha\beta}$ vanish outside the time interval of GW production, namely
\begin{align}
\label{ABK}
\pounds_T T_{\alpha\beta}[\eta < \eta_{\text{i}}] = \pounds_T T_{\alpha\beta}[\eta > \eta_{\text{f}}] = 0 .
\end{align}
In contrast, we appeared to have made stronger assumptions (equations \eqref{ShearStressZero} and \eqref{MassQuadrupoleDotZero}), but without appealing to the symmetries enjoyed by de Sitter spacetime. One reason is that we wish to analyze the more general case of spatially flat FLRW-like geometries in arbitrary dimensions, where $T^\mu \partial_\mu$ in eq. \eqref{KillingVector} is no longer a Killing vector. Moreover, we also wish to point out that eq. \eqref{ABK} may lead to a potential pathology if one does not assume $T_{ij}[\eta > \eta_\text{f}] = 0$ in eq. \eqref{ShearStressZero}. This is because, a direct calculation would show that
\begin{align}
\label{LieDerivative}
\pounds_T T_{\alpha\beta}
&= -H \left( \partial_\sigma (x^\sigma T_{\alpha\beta}) - (d-2) T_{\alpha\beta} \right) .
\end{align}
For later purposes, we also record the alternate form
\begin{align}
\label{LieDerivative_v2}
\pounds_T T_{\alpha\beta}
&= -H \left\{ \frac{\partial_0 \left( a^{d-1} T_{\alpha\beta} \right) - Had (a^{d-1} T_{\alpha\beta})}{-Ha^d} + \partial_l \left( x^l T_{\alpha\beta}\right) - (d-2) T_{\alpha\beta} \right\} .
\end{align}
Integrating eq. \eqref{ABK} over space, using the form of the Lie derivative in eq. \eqref{LieDerivative}, then leads us to the following differential equation, valid for $\eta > \eta_{\text{f}}$,
\begin{align}
0 = \partial_0 \left(\frac{\widehat{T}_{\alpha\beta}[\eta]}{a[\eta]}\right) + H (d-2) \widehat{T}_{\alpha\beta}[\eta], \qquad\text{ where }\qquad
\widehat{T}_{\alpha\beta}[\eta] \equiv \int_{\mathbb{R}^{d-1}} \dd^{d-1}\vec{x} T_{\alpha\beta}[\eta,\vec{x}] .
\end{align}
The solution for $\eta \geq \eta_\text{f}$ is
\begin{align}
a[\eta]^{d-2} \widehat{T}_{\alpha\beta}[\eta] = \frac{a[\eta]}{a[\eta_\text{f}]} a[\eta_\text{f}]^{d-2} \widehat{T}_{\alpha\beta}[\eta_\text{f}] ,
\end{align}
indicating that the tail integral in eq. \eqref{dS_Even_Tails} would yield a divergent $\overline{\chi}_{ij}$ -- this includes its gauge-invariant part $D_{ij}$ -- in the asymptotic future $\eta_r \to 0^-$ unless $\widehat{T}_{ij}[\eta_\text{f}] = 0$. In particular, there is a ``log-divergence" of the form
\begin{align}
\overline{\chi}_{ij}^\text{(tail)}[\eta-r \to 0^-,\vec{x}]
\propto \lim_{\eta_> \to 0^-} \int_{\eta_{\text{f}}}^{\eta_>} \dd\eta \int_{\mathbb{R}^{d-1}} \dd^{d-1} \vec{x} a^{d-2} T_{ab}[\eta,\vec{x}]
\propto \lim_{\eta_> \to 0^-} \ln[\eta_>/\eta_\text{f}] .
\end{align}
{\bf Remark II} \qquad In eq. \eqref{MassQuadrupoleDotZero} we have chosen to assume, from the outset, that the mass quadrupole moments do not evolve outside the time interval when GWs are generated. It is possible to provide partial justifications of this assumption in de Sitter spacetime.

Let us continue to assume eq. \eqref{ShearStressZero} (and thus eq. \eqref{PressureQuadrupoleZero}) but now keep the time derivative of the mass quadrupole moment in eq. \eqref{Conservation_dS} when evaluating eq. \eqref{dS_Even_Tails}. At late times,
\begin{align}
\label{TailIntegral}
\overline{\chi}_{ij}^{\text{(tail)}}[\eta_r > \eta_\text{f},\vec{x}]
&= -8\pi \GN \frac{H^{d-2}}{(2\pi)^{\frac{d-2}{2}}} \frac{(d-2)!}{2^{\frac{d}{2}} \left(\frac{d-2}{2}\right)!}  
	\left[ \frac{\partial_0 Q_{ab}[\eta']}{a[\eta']} - 2 H Q_{ab}[\eta'] \right]_{\eta'=\eta_\text{i}}^{\eta'=\eta_\text{f}} .
\end{align}
Had we extended the upper limit of integration in eq. \eqref{TailIntegral} from $\eta_{\text{f}}$ to $\eta_> > \eta_{\text{f}}$, the result should not change because $T_{ij}$ is zero between $\eta_{\text{f}} \leq \eta \leq \eta_>$. That means $a[\eta]^{-1} \partial_0 Q_{ij}[\eta] -2H Q_{ij}[\eta] = a[\eta_\text{f}]^{-1} \partial_0 Q_{ij}[\eta_\text{f}] -2H Q_{ij}[\eta_\text{f}] (=$ constant), whose solution is
\begin{align}
\label{MassQuadrupole_LateEvolution}
Q_{ij}[\eta \geq \eta_{\text{f}}] 
= Q_{ij}[\eta_{\text{f}}] + \frac{1}{2 H a[\eta_\text{f}]} \frac{\partial Q_{ij}[\eta_{\text{f}}]}{\partial \eta_{\text{f}}} \left( \left(\frac{a[\eta]}{a[\eta_{\text{f}}]}\right)^2 - 1 \right) .
\end{align}
However, as $\eta \to 0$ (i.e., towards asymptotic future) it appears the physical quadrupole moment itself will blow up unless $\dot{Q}_{ij}[\eta \geq \eta_{\text{f}}]=0$. It thus appears reasonable to demand the time derivative of the mass quadrupole moment to vanish at late times,
\begin{align}
\label{MassQuadrupole_LateTime}
\dot{Q}_{ij}[\eta \geq \eta_\text{f}] = 0 .
\end{align}
{\bf Remark III} \qquad If we follow \cite{Ashtekar:2015lxa} and assume the static condition in eq. \eqref{ABK}, it is then possible to justify why both the mass and pressure quadrupole moments stay constant outside the interval of active GW production, $\eta_\text{i} \leq \eta' \leq \eta_\text{f}$. To see this we multiply $x^i x^j$ to the $00$ component and spatial-trace of the static condition in eq. \eqref{ABK}, through the form of the Lie derivative in eq. \eqref{LieDerivative_v2}. Proceeding to integrate the resulting expressions over all space while employing the definitions of the mass and pressure quadrupole moments in equations \eqref{MassQuadrupole} and \eqref{PressureQuadrupole}, we obtain
\begin{align}
\partial_0 Q^{ij} - H a d Q^{ij} - H a \int_{\mathbb{R}^{d-1}} \dd^{d-1}\vec{x} ( x^i x^j ) \partial_l (x^l T_{00}) a^{d-1}
		&= -(d-2) H a Q^{ij} , \\
\partial_0 P^{ij} - H a d P^{ij} - H a \int_{\mathbb{R}^{d-1}} \dd^{d-1}\vec{x} ( x^i x^j ) \partial_l (x^l \delta^{mn} T_{mn}) a^{d-1}
		&= -(d-2) H a P^{ij} .
\end{align}
Integrating-by-parts the third terms from the left, exploiting the physically localized nature of the source to set surface terms to zero, and again recalling \eqref{MassQuadrupole} and \eqref{PressureQuadrupole}, we arrive at the already advertised assertion:\footnote{We believe eq. \eqref{MassPressureQuadrupolesDotZero} is consistent with eq. (4.26) in \cite{Ashtekar:2015lxa}. Note that, their $Q_{ab}^{(\rho)}$ is our $Q_{ij}$ and their $Q_{ab}^{(p)}$ is our $P_{ij}$; according to them, $\pounds_T Q_{ab}^{(\rho,p)} = \partial_t Q_{ab}^{(\rho,p)} -2 H Q_{ab}^{(\rho,p)}$, where $t$ is observer time defined by $\dd t = a[\eta] \dd\eta$. However, it is unclear why they treat $Q_{ab}^{(\rho)}$ and $Q_{ab}^{(p)}$ as tensors, as indicated by their taking of the quadrupoles' Lie derivatives. This is most likely why they go on to state that eq. \eqref{ABK} ``{\it \dots does not imply that quadrupoles are left invariant by the flow generated by $T^a$}". It is easy to misconstrue this, as if it were claiming that $Q_{ab}^{(\rho)}$ and $Q_{ab}^{(p)}$ are no longer time independent in spite of eq. \eqref{ABK}. Instead, we do not believe the mass and pressure quadrupole moments are coordinate tensors; they do transform covariantly under global rotations of the spatial coordinates $\vec{x}$, but are otherwise merely one-parameter (i.e., time-dependent) objects describing aspects of the GW source(s)' internal structure.}
\begin{align}
\label{MassPressureQuadrupolesDotZero}
\dot{Q}_{ij} = \dot{P}_{ij} = 0 \qquad\qquad \text{ whenever } \qquad\qquad \pounds_T T_{\alpha\beta} = 0 .
\end{align}

\section{Linear GW Memory Effects In A 4D Spatially Flat Cosmology}
\label{Section_MemoryEffect_4DCosmology}
In this section, we elaborate on how the spatially flat FLRW results in \cite{Chu:2015yua} lead to linear GW memory effects in an expanding 4D universe like ours. Notationally, we will say that the TT GWs split into
\begin{align}
D_{ij} = D_{ij}^{(\gamma)} + D_{ij}^{\text{(tail)}} ,
\end{align}
where $D_{ij}^{(\gamma)}$ are the gravitons that travel on the light cone and $D_{ij}^{\text{(tail)}}$ are the ones traveling inside the null cone.

{\bf On the null cone} \qquad The portion of the GW signal that transmits information on the light cone was shown in \cite{Chu:2015yua} to take a form very similar to its counterpart in Minkowski spacetime:
\begin{align}
\label{4D_GW_LightCone}
D_{ij}^{(\gamma)}[\eta,\vec{x}] = -4 \GN \int_{\mathbb{R}^3} \dd^3 \vec{x}' a[\eta_r]^3 \frac{T_{\widehat{i}\widehat{j}}^\text{TT}[\eta_r,\vec{x}']}{a[\eta] |\vec{x}-\vec{x}'|},
\qquad \eta_r \equiv \eta - |\vec{x}-\vec{x}'| .
\end{align}
This holds for any cosmic history $a[\eta]$; and the flat spacetime limit is recovered by setting $a \to 1$. Since $T_{\widehat{i}\widehat{j}} = a^{-2} T_{ij}$ is the physical shear-stress density of the isolated source, provided its internal dynamics does not span cosmological timescales, we wish to assert here that all the known linear GW memory effects in 4D asymptotically flat spacetimes would carry over to the asymptotically-cosmological case at hand. For instance, $T_{\widehat{i}\widehat{j}}$ could describe two gravitationally unbound compact bodies flying past each other, giving rise to a ``burst" of GWs \cite{BraginskyThorne}. The primary difference is that GWs traveling on the null cone in a 4D spatially flat universe will now be further diluted by cosmic expansion due to the scale factor in the $1/(a[\eta]|\vec{x}-\vec{x}'|)$ fall-off.

{\it Far zone} \qquad In the far zone, we replace $|\vec{x}-\vec{x}'| \to r \equiv |\vec{x}|$ when computing the GW. This then allows us to employ eq. \eqref{Conservation_III} in eq. \eqref{4D_GW_LightCone}. With $\eta_r \approx \eta - r$,
\begin{align}
D_{ij}^{(\gamma)}[\eta,\vec{x}] 
&\approx -\frac{2 \GN}{a[\eta] a[\eta_r] r} \partial_0 \left\{
\frac{\partial_0 Q_{ij}^\text{TT}[\eta_r]}{a[\eta_r]} - \frac{\dot{a}[\eta_r]}{a[\eta_r]^2} \left(2Q_{ij}^\text{TT}[\eta_r] - P_{ij}^\text{TT}[\eta_r]\right) \right\} .
\end{align}
{\bf Inside the null cone} \qquad Unlike its null cone counterpart, the detailed structure of the TT GW tail can only be determined after its relevant wave equation is solved.

{\it Radiation domination} \qquad In \cite{Chu:2015yua} we showed, in particular, that there are no radiative GW tails in a radiation dominated universe ($a[\eta] = \eta/\eta_0$) because the background Ricci scalar is zero. Therefore linear GWs in such a cosmology, including potential memory effects, are fully captured by eq. \eqref{4D_GW_LightCone}.

{\it Matter domination} \qquad The scale factor for a matter dominated universe is $a[\eta] = (\eta/\eta_0)^2$, and its TT GW tail is
\begin{align}
D_{ij}^{\text{(tail)}}[\eta,\vec{x}] &= -\frac{4 \GN}{\eta_0^2 a[\eta]^{\frac{3}{2}}} \int_{\mathbb{R}^3} \dd^3 \vec{x}' \int_0^{\eta_r} \dd\eta' a[\eta']^{\frac{5}{2}} T_{\widehat{i}\widehat{j}}^\text{TT}[\eta',\vec{x}'] , \\
\eta_r &\equiv \eta-|\vec{x}-\vec{x}'|-0^+ . \nonumber 
\end{align}
As already alluded to in \cite{Chu:2015yua}, after the source has ceased ($\eta_r > \eta_\text{f}$), the GW tail becomes space-independent, and like its de Sitter counterpart, does not fall off with increasing distance from the source itself. Assuming the vanishing of shear-stress outside the time interval of active GW genesis (i.e., eq. \eqref{ShearStressZero}), we have
\begin{align}
\label{4D_GW_Tail_MatterDom}
D_{ij}^{\text{(tail)}}[\eta - r > \eta_\text{f},\vec{x}] 
&= -\frac{4 \GN}{\eta_0^2 a[\eta]^{\frac{3}{2}}} \int_{\mathbb{R}^3} \dd^3 \vec{x}' \int_{\eta_{\text{i}}}^{\eta_{\text{f}}} \dd\eta' a[\eta']^{\frac{5}{2}} T_{\widehat{i}\widehat{j}}^\text{TT}[\eta',\vec{x}'] .
\end{align}
Moreover, since the scale factor changes appreciably only over cosmological timescales ($\eta_0 \sim 10$ Gyr), note that over the timescales of human GW experiments ($\lesssim$ decades) we may, as a good approximation, regard the tail in eq. \eqref{4D_GW_Tail_MatterDom} as a spacetime constant (for $\eta_r > \eta_\text{f}$). Notice too, because the power of the scale factor is $5/2$, the integrand in eq. \eqref{4D_GW_Tail_MatterDom} cannot be converted into a total time derivative involving the mass and pressure quadrupole moments, unlike the de Sitter case; recall eq. \eqref{Conservation_III}.

To sum: there is a tail induced linear GW memory effect in a 4D spatially flat matter dominated FLRW universe. Its amplitude does not depend on space, but does decay over cosmological timescales in an expanding cosmology.

{\it de Sitter} \qquad Since we have already elaborated upon the tail induced linear GW memory effect in even dimensional $d \geq 4$ de Sitter, we close this section by comparing the null cone (aka in the GR literature as the ``direct part") versus the tail part of the TT GW in 4D. For simplicity, we shall take the far zone limit in the following equations. Expressed in terms of the mass and pressure quadrupole moments, the direct part reads
\begin{align}
\label{dS_4D_Direct_FarZone}
D_{ij}^{(\gamma)}[\eta,\vec{x}] \approx -\frac{2\GN}{a[\eta] a[\eta_r] |\vec{x}|} \left\{ \frac{\ddot{Q}_{ij}^\text{TT}[\eta_r]}{a[\eta_r]} - H \left( 3 \dot{Q}_{ij}^\text{TT}[\eta_r] - \dot{P}_{ij}^\text{TT}[\eta_r] \right) \right\},
\qquad \eta_r \approx \eta - |\vec{x}| .
\end{align}
On the other hand, the linear TT GW tail is, according to eq. \eqref{dS_Even_Tails_QuadrupoleMoments},
\begin{align}
\label{dS_4D_Tail_FarZone}
D_{ij}^{\text{(tail)}}[\eta,\vec{x}] 
\approx -2 H^2 \GN 
\left[ \frac{\dot{Q}_{ij}^\text{TT}[\eta']}{a[\eta']} - H \left( 2 Q_{ij}^\text{TT}[\eta'] - P_{ij}^\text{TT}[\eta'] \right) \right]_{\eta' = -\infty}^{\eta' = \eta_r} .
\end{align}
The flat spacetime limit can be obtained by setting $H \to 0$, which in turn implies replacing $a \to 1$ and all conformal-time derivatives with Minkowski-time derivatives. Observe that the second and third terms on the right hand sides of equations \eqref{dS_4D_Direct_FarZone} and \eqref{dS_4D_Tail_FarZone} are sub-dominant to the first terms in this limit, since they are multiplied by an additional power of $H$. Moreover, because of the overall $H^2$ multiplicative factor, the tail term of eq. \eqref{dS_4D_Tail_FarZone} tends smoothly to zero as $H$ (and hence the cosmological constant $\Lambda$) vanishes. As already estimated in \cite{Chu:2015yua} -- and re-confirmed by dividing the right hand sides of \eqref{dS_4D_Tail_FarZone} and 
\eqref{dS_4D_Direct_FarZone} while keeping only their first terms -- the ratio of the GW tail amplitude to that of the direct part is roughly $|D_{ij}^{(\text{tail})}/D_{ij}^{(\gamma)}| \sim (H \cdot \Delta t)(H \cdot a[\eta] |\vec{x}|)$, where $\Delta t$ is the physical characteristic time scale associated with the GW source and $a|\vec{x}|$ is the physical observer-source distance at the observer's time $\eta$. In other words, unless the GW generation process takes place at cosmological distances from the observer and over cosmological timescales, the tail signal is highly suppressed whenever the direct signal is simultaneously present.

However, if we assume that the shear-stress density becomes negligible and the quadrupole moments settle to static values outside the time interval of active GW production -- i.e., equations \eqref{ShearStressZero}, \eqref{PressureQuadrupoleZero} and \eqref{MassQuadrupoleDotZero} -- then, in the late time limit, $\eta_r > \eta_\text{f}$, the direct GW contribution from eq. \eqref{dS_4D_Direct_FarZone} vanishes and what remains is the tail part from eq. \eqref{dS_4D_Tail_FarZone}.
\begin{align}
D_{ij}[\eta_r > \eta_\text{f}] 
= D_{ij}^{\text{(tail)}}[\eta_r > \eta_\text{f}] 
\approx 4 H^3 \GN \left( Q_{ij}^\text{TT}[\eta_\text{f}] - Q_{ij}^\text{TT}[\eta_\text{i}] \right).
\end{align}
%In the present dark-energy dominated epoch of our universe, with $H \sim 10^{-32}$ eV, one can perform a crude estimate of the fractional distortion (eq. \eqref{FractionalDistortion}) produced by the GW tail of the recently detected binary black hole (BBH) merger, whose signal was named GW150914 \cite{Abbott:2016blz}. With $\GN \sim 1/\mpl^2 \sim 6.7 \times 10^{-39}/\text{GeV}^2$; approximating the BBH quadrupole moment $Q_{ij}$ as the total mass $(\sim 100 M_\odot)$ times the square of the characteristic size of the merged BHs (i.e., its Schwarzschild radius), 
%\begin{align}
% D_{ij}^{\text{(tail)}} \sim (H^2 \GN) \cdot (H \times 100 M_\odot) (100 \GN M_\odot)^2 \sim 10^{-60} .
%\end{align}

\section{Summary and Future Directions}
\label{Section_SummaryFutureDirections}
\begin{figure}
\begin{center}
\includegraphics[width=4in]{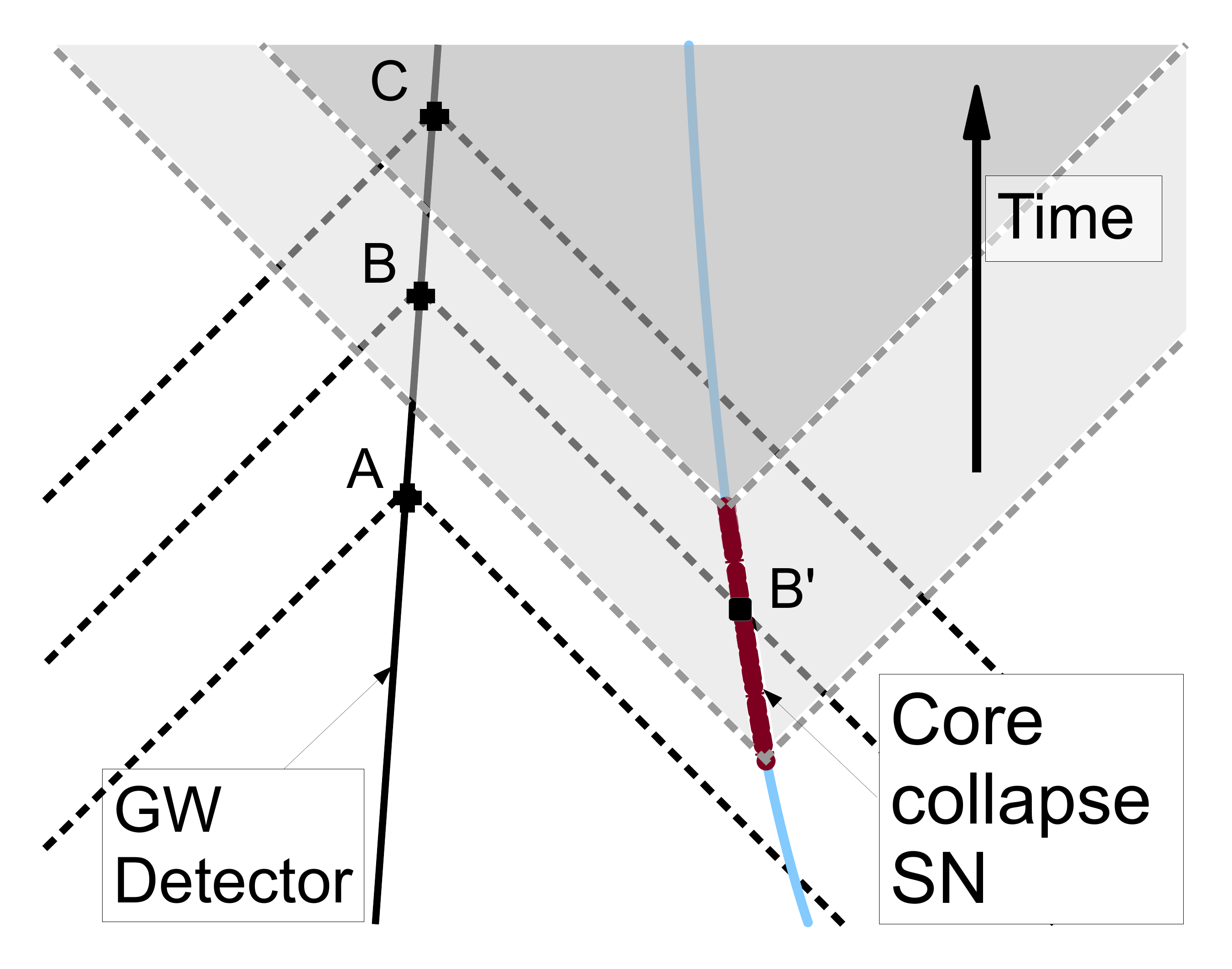}
\caption{(Figure borrowed from \cite{Chu:2015yua}.) This is a spacetime diagram depicting the core collapse of a massive star, which then goes supernova (right world line). The dashed-dotted segment of the right world line denotes the full duration during which TT GWs are produced, corresponding to $\eta_\text{i} \leq \eta \leq \eta_\text{f}$ in the main text. For $\eta \leq \eta_\text{i}$, the collapse has not started; afterwards, $\eta \geq \eta_\text{f}$, the system has settled down completely. (We assume equations \eqref{ShearStressZero}, \eqref{PressureQuadrupoleZero} and \eqref{MassQuadrupoleDotZero}.) The TT GWs are heard by a distant detector (left world line). In this paper the background geometry is either dS$_{4+2n}$, a spatially flat 4D FLRW radiation dominated or matter dominated universe. The black dashed lines emanating from the worldline of the GW detector are the past light cones of events $A$, $B$ and $C$. The bottom pair of light gray dashed lines emanating from the right world line is the forward light cone of the starting point of the stellar collapse; the top pair is that of the ending point. The light gray shaded region of spacetime is filled with TT GWs propagating both on and inside the light cone. The darker-gray region of spacetime is filled with TT GW tails only, whose detailed properties depends in principle on the entire history of the source (i.e., the dashed-dotted segment). A detector that was operational from $A$ through $C$ would sense a permanent change in $D_{ij} \equiv \chi_{ij}^\text{TT}$ if the TT GW tails in this darker-gray region were spacetime constant. This happens in dS$_{4+2n}$ (eq. \eqref{dS_Memory}) and approximately so in 4D matter dominated universes (eq. \eqref{4D_GW_Tail_MatterDom}) -- but not in 4D radiation dominated ones, because there are no TT GW tails there.}
\label{TTGWFigure}
\end{center}
\end{figure}
In this paper, we have improved upon and extended the 4D cosmological results of \cite{Chu:2015yua}. Following Ashtekar, Bonga, and Kesavan \cite{Ashtekar:2015lxa} and Date and Hoque \cite{Date:2015kma}, we have now expressed the TT GW tails in even $(d \geq 4)$-dimensional de Sitter spacetime directly in terms of the mass and pressure quadrupole moments of the source. Additionally, we have obtained the solutions to Einstein's equations, with a positive cosmological constant, linearized about a background de Sitter spacetime of dimensions greater or equal to four. In particular, we have shown that the 4D inside-the-null-cone linear GW memory effect found in \cite{Chu:2015yua}, due to the spacetime constant tail of the massless TT graviton (and scalar) Green's function, really extends to all higher dimensional dS$_{4+2n}$. We also suggested that the known linear GW memory effects in asymptotically flat 4D spacetimes will carry over to the asymptotically spatially flat FLRW case, with the additional feature that the TT GW amplitude will be diluted by cosmic expansion. Even though radiation dominated universes exhibit no TT GW tails, matter dominated ones do -- and in fact, yield approximately spacetime constant GW tails that, like their de Sitter cousins, induce a linear memory effect. In Fig. \eqref{TTGWFigure} we summarize/illustrate our findings in sections \eqref{Section_MemoryEffect_dS} and \eqref{Section_MemoryEffect_4DCosmology}.

On a Minkowski background the energy-momentum-shear-stress (pseudo-)tensor of the GWs themselves source a nonlinear (aka ``Christodoulou") memory effect \cite{Christodoulou:1991cr}; see also \cite{BraginskyThorne} and \cite{Thorne1992}. How this will generalize to a de Sitter background geometry is, as far as we are aware, a wide open question. On the other hand, that we have been able to successfully utilize Nariai's ansatz \cite{Nariai} to solve Einstein's equations linearized on dS$_{d \geq 4}$ suggests it may be worthwhile to try solving them for a background spatially flat FLRW geometry with a more general equation-of-state $w$. These results could not only provide further insight into potential GW memory effects, they could lead to new perspectives on the study of large scale structure in our universe. Finally, even though \cite{Chu:2013hra} has obtained the $d$-dimensional minimally coupled massless scalar de Sitter Green's function from its $(d+1)$-dimensional Minkowski counterpart, the question of how to do so for the full graviton Green's function -- the collection of $\mathcal{G}$s in equations \eqref{dS_PseudoTrace_Solution}, \eqref{dS_Vector_Solution}, and \eqref{dS_Tensor_Solution} -- is still unanswered.

\section{Acknowledgments}

I thank members of the cosmology group here at the University of Minnesota Duluth for useful discussions, in particular: Hadi Papei, Vitaly Vanchurin and Eric West. I am grateful to Dai De-Chang for referring me to Nariai's work \cite{Nariai}, and to Tanmay Vachaspati for valuable advice. I have also benefited from e-mail exchanges with Abhay Ashtekar, Ghanashyam Date, Alex Kehagias and Antonio Riotto. A portion of this work was carried out at the Tavern On The Hill; I wish to thank the wonderful staff there -- Jamie Lakatos, Justin Richard, Michael Simon, Katie Trangsrud, Nicholas Weber, and many more -- for providing a welcoming and stimulating environment. Much of the analytic work here was performed with the aid of {\sf Mathematica} \cite{Mathematica} and the tensor package {\sf xAct} \cite{xAct}. 

\appendix
\section{Geodesic spatial distance between a pair of test masses}
\label{Section_GeodesicDistance}
The main goal of this section is the derivation of eq. \eqref{FractionalDistortion}. We wish to calculate the geodesic spatial distance between two test masses, at a fixed time $\eta$, when one is located (in space) at $\vec{X}$ and the other at $\vec{X}'$. (These test masses are not assumed to follow geodesics.) We note that, if the spacetime geometry is described by eq. \eqref{PerturbedFLRW}, the induced metric on a constant$-\eta$ hypersurface is $\dd\vec{\ell}^2 \equiv -a[\eta]^2 (\delta_{ij} - \chi_{ij}[\eta,\vec{x}])\dd x^i \dd x^j$. The square of the geodesic spatial distance between $\vec{X}$ and $\vec{X}'$ is given by the integral
\begin{align}
\label{SpatialSyngeWorldFunction}
L[\eta]^2 
= a[\eta]^2 \int_0^1 \left( \delta_{ij} - \chi_{ij}\left[ \eta,\vec{Y}[\lambda] \right] \right) \frac{\dd Y^i}{\dd \lambda} \frac{\dd Y^j}{\dd \lambda} \dd\lambda,
\end{align}
where $Y^i$ obeys the geodesic equation
\begin{align}
\label{SpatialGeodesicEquation}
\left. \frac{\dd^2 Y^i}{\dd \lambda^2} 
+ \Gamma^i_{\phantom{i}jk}\left[g_{ab}\right] \frac{\dd Y^j}{\dd \lambda} \frac{\dd Y^k}{\dd \lambda} \right\vert_{g_{ab} \equiv \delta_{ab} - \chi_{ab}} = 0 ,
\end{align}
and the boundary conditions
\begin{align}
Y^i[\lambda = 0] = X'^i \qquad \text{ and } \qquad Y^i[\lambda = 1] = X^i .
\end{align}
Now, if there were no perturbations, i.e., $\chi_{ij} = 0$, then the result is simply (up to the overall factor $a^2$) the familiar one in Euclidean space
\begin{align}
L_0[\eta]^2 = a[\eta]^2 \int_0^1 \delta_{ij} \frac{\dd X_0^i}{\dd \lambda} \frac{\dd X_0^j}{\dd \lambda} \dd\lambda = a[\eta]^2 |\vec{X}-\vec{X}'|^2 ,
\end{align}
where
\begin{align}
\label{StraightLine}
\vec{X}_0[\lambda] \equiv \vec{X}' + \lambda (\vec{X}-\vec{X}') ,
\end{align} 
a straight line joining $\vec{X}'$ to $\vec{X}$ as $\lambda$ runs from $0$ to $1$. Notice the solution $Y^i$ of the geodesic equation in eq. \eqref{SpatialGeodesicEquation} has to deviate from a straight line in Euclidean space by terms of $\mathcal{O}[\chi_{ij}]$ and higher. Because eq. \eqref{SpatialSyngeWorldFunction} with $\chi_{ij}$ set to zero is also the variational principle that leads to the geodesic equation in Euclidean space, that means if we evaluate eq. \eqref{SpatialSyngeWorldFunction} by replacing all the $Y^i$ with $X_0^i$ in eq. \eqref{StraightLine}, the error incurred begins at order $\mathcal{O}[(\chi_{ij})^2]$. Doing so leads us to
\begin{align}
\label{GeodesicSpatialDistance}
L[\eta]
= a[\eta] |\vec{X}-\vec{X}'| \left( 1 - \frac{\widehat{n}^i \widehat{n}^j}{2} \int_{0}^{1} \chi_{ij}\left[ \eta,\vec{X}_0[\lambda] \right] \dd\lambda \right)
+ \mathcal{O}\left[ (\chi_{ij})^2 \right] ,
\end{align}
with $\widehat{n} \equiv (\vec{X}-\vec{X}')/|\vec{X}-\vec{X}'|$. From this result, we see that the change in geodesic spatial distance due to the presence of the metric perturbation is
\begin{align}
\label{DeviationOfGeodesicDistance}
\delta L[\eta]
= - a[\eta] |\vec{X}-\vec{X}'| \frac{\widehat{n}^i \widehat{n}^j}{2} \int_{0}^{1} \chi_{ij}\left[ \eta,\vec{X}_0[\lambda] \right] \dd\lambda
+ \mathcal{O}\left[ (\chi_{ij})^2 \right] .
\end{align}
Dividing the result in eq. \eqref{DeviationOfGeodesicDistance} with that in eq. \eqref{GeodesicSpatialDistance}, and keeping only terms linear in $\chi_{ij}$, establishes eq. \eqref{FractionalDistortion}.

\section{Solution to a space-translation-invariant PDE via dimension reduction}
\label{Section_PDESolution}
The central equations of this paper are \eqref{dS_PseudoTrace_IIofII}, \eqref{dS_Vector_IIofII} and \eqref{dS_Tensor_IIofII}, taking the form
\begin{align}
\label{G_MotivatingPDE}
\left(\partial_x^2 + U[\eta]\right) \left( a[\eta]^{\frac{d-2}{2}} \psi[\eta,\vec{x}] \right) = -16 \pi \GN a[\eta]^{\frac{d-2}{2}} S[\eta,\vec{x}] ,
\end{align}
where $U$ is a time dependent potential, $\psi$ is some component(s) of the barred graviton, and $S$ is some component(s) of $T_{\mu\nu}$. In this section we wish to solve the associated symmetric (retarded plus advanced) Green's function equation:
\begin{align}
\label{G_MasterPDE}
\left(\partial_x^2 + U[\eta]\right)\widehat{G}_d[\eta,\eta';R \equiv |\vec{x}-\vec{x}'|] 
&= \left(\partial_{x'}^2 + U[\eta']\right)\widehat{G}_d[\eta,\eta';R \equiv |\vec{x}-\vec{x}'|] \nonumber\\
&= 2 \delta[\eta-\eta'] \delta^{(d-1)}[\vec{x}-\vec{x}'] ,
\end{align}
where
\begin{align}
\partial_x^2 \equiv \eta^{\mu\nu} \frac{\partial}{\partial x^\mu} \frac{\partial}{\partial x^\nu}, \qquad\qquad
\partial_{x'}^2 \equiv \eta^{\mu\nu} \frac{\partial}{\partial x'^\mu} \frac{\partial}{\partial x'^\nu} ,
\end{align}
with $x^\mu \equiv (\eta,\vec{x})$, $x'^\mu \equiv (\eta',\vec{x}')$, and $|\vec{x}-\vec{x}'|$ is the Euclidean distance between $\vec{x}$ and $\vec{x}'$. We will obtain explicit solutions for the case where
\begin{align}
\label{PowerLawPotential}
U[\eta] \equiv -\frac{p}{\eta^2}, \qquad \qquad p \in \mathbb{R} ,
\end{align}
but we will begin with the general $U[\eta]$. For the de Sitter analysis at hand, equations \eqref{dS_PseudoTrace_IIofII}, \eqref{dS_Vector_IIofII} and \eqref{dS_Tensor_IIofII} tell us the relevant $p$ are $(d-6)(d-4)/4$, $(d-4)(d-2)/4$, and $d(d-2)/4$.

Because the potential $U$ is time dependent but space independent, we expect the Green's function to reflect the space-translation-invariance of the differential operator. In particular, as long as space is assumed to be infinite, we may employ the integral relationship between the Green's functions in $d$ and $d+2$ dimensions:
\begin{align}
\label{G_Recursion_PlaneSource}
\widehat{G}_{d}\left[ \eta,\eta';R \right] 
= \int_{\mathbb{R}^2} \dd^2 \vec{x}_\perp \widehat{G}_{d+2}\left[ \eta,\eta';\sqrt{R^2 + \vec{x}_\perp^2} \right] ,
\end{align}
where $\vec{x}_\perp = (x^d,x^{d+1})$ are the two extra spatial dimensions. That the $d-$dimensional version of eq. \eqref{G_MasterPDE} is satisfied by eq. \eqref{G_Recursion_PlaneSource} -- provided the $(d+2)-$version holds -- can be verified by applying $\Box_{d} + U$ (with respect to either the $d$-dimensional unprimed or primed variables) on both sides, where $\Box_{d} \equiv \partial_\mu \partial^\mu$ or $\Box_{d} \equiv \partial_{\mu'} \partial^{\mu'}$. If $\vec{\nabla}_\perp^2 = (\partial/\partial x^d)^2 + (\partial/\partial x^{d+1})^2$ (or $\vec{\nabla}_\perp^2 = (\partial/\partial x'^d)^2 + (\partial/\partial x'^{d+1})^2$) is the Laplacian with respect to $\vec{x}_\perp$, we may then utilize $\Box_{d+2} = \Box_{d} - \vec{\nabla}_\perp^2$, and deduce
\begin{align}
(\Box_d + U)\widehat{G}_{d}\left[ \eta,\eta';R \right] 
&= \int_{\mathbb{R}^2} \dd^2 \vec{x}_\perp \left( \Box_{d+2} + U \right)\widehat{G}_{d+2}\left[ \eta,\eta';\sqrt{R^2 + \vec{x}_\perp^2} \right] \nonumber\\
&\qquad 
+ \int_{\mathbb{R}^2} \dd^2 \vec{x}_\perp \vec{\nabla}_\perp^2 \widehat{G}_{d+2}\left[ \eta,\eta';\sqrt{R^2 + \vec{x}_\perp^2} \right] .
\end{align}
Using the $(d+2)$-dimensional version of eq. \eqref{G_MasterPDE} on the right hand side yields $2 \delta[\eta-\eta'] \delta^{(d-1)}[\vec{x}-\vec{x}']$ as desired, provided the second line is zero. The latter is true because the integrand is a total spatial derivative with respect to $\vec{x}_\perp$, and thus can be converted into a surface integral lying well outside the light cone of $x^\mu_{d+2} \equiv (\eta,\vec{x},\vec{0}_\perp)$. By causality, the symmetric Green's function $\widehat{G}_{d+2}$ has to vanish there. Alternatively, eq. \eqref{G_Recursion_PlaneSource} can also be proven by going to Fourier space.

By writing $\rho \equiv \sqrt{R^2 + \vec{x}_\perp^2}$, eq. \eqref{G_Recursion_PlaneSource} now becomes
\begin{align}
\widehat{G}_{d}\left[ \eta,\eta';R \right] 
= 2\pi \int_{R}^{\infty} \dd \rho \cdot \rho \ \widehat{G}_{d+2}\left[ \eta,\eta';\rho \right] ,
\end{align}
and by differentiating both sides with respect to $R$, we obtain the following recursion relation between the symmetric Green's function in $d$ and $d+2$ dimensions:
\begin{align}
\label{G_Recursion}
\widehat{G}_{d+2}[\eta,\eta';\sbar] = \frac{1}{2\pi} \frac{\partial \widehat{G}_d[\eta,\eta';\sbar]}{\partial \sbar} .
\end{align}
Here, we have re-expressed the derivative with respect to the Euclidean distance $R$ into one with respect to Synge's world function in flat spacetime,
\begin{align}
\label{SyngeWorldFunctionMinkowski}
\sbar \equiv \frac{1}{2}\left( (\eta-\eta')^2 - R^2 \right) = \frac{1}{2}\eta_{\mu\nu} (x-x')^\mu (x-x')^\nu .
\end{align}
We see that once the $d=2,3$ results are known, the higher dimensional ones follow from differentiation.
\begin{align}
\label{G_Recursion_Even}
\widehat{G}_{\text{even $d$}}[\eta,\eta';\sbar] 
	&= \frac{1}{(2\pi)^{\frac{d-2}{2}}} 
		\left(\frac{\partial}{\partial \sbar}\right)^{\frac{d-2}{2}} \widehat{G}_2[\eta,\eta';\sbar] , \\
\label{G_Recursion_Odd}
\widehat{G}_{\text{odd $d$}}[\eta,\eta';\sbar] 
	&= \frac{1}{(2\pi)^{\frac{d-3}{2}}} 
		\left(\frac{\partial}{\partial \sbar}\right)^{\frac{d-3}{2}} \widehat{G}_3[\eta,\eta';\sbar] .
\end{align}
Note that the retarded Green's functions $\widehat{G}^+$ follow from their symmetric ones $\widehat{G}$ by multiplying the latter by $\Theta[\eta-\eta']$, where $\Theta[z > 0] = 1$ and $\Theta[z<0] = 0$,
\begin{align}
\widehat{G}^+\left[ \eta,\vec{x};\eta',\vec{x}' \right] = \Theta[\eta-\eta'] \widehat{G}\left[ \eta,\eta';\sbar \right] .
\end{align}

\subsection{Even Dimensions}

In two dimensions we first postulate the following ansatz for the symmetric Green's function:
\begin{align}
\label{G_2DAnsatz}
\widehat{G}_2[\eta,\eta';\sbar] = \frac{\Theta[\sbar]}{2} J[\eta,\eta';\sbar] .
\end{align}
Inserting eq. \eqref{G_2DAnsatz} into eq. \eqref{G_MasterPDE}, and noting that $\Theta[\sbar]/2$ is the symmetric minimally coupled massless scalar 2D Green's function,
\begin{align}
\partial^2 \left( \frac{\Theta[\sbar]}{2} \right) = 2 \delta^{(2)}[x-x'] ,
\end{align}
one would deduce
\begin{align}
\label{G_2DAnsatz_MainStep}
\left(\partial^2 + U\right)\widehat{G}_2[\eta,\eta';\sbar] 
&= 2 \delta^{(2)}[x-x'] J[x=x'] \\
&\qquad
+ \sigma_s \delta[\sbar] (x-x')^\alpha \partial_\alpha J[\sbar=0] 
+ \frac{\Theta[\sbar]}{2} \left(\partial^2 + U\right) J , \nonumber
\end{align}
where $\sigma_s = +1$ if $\partial^2$ was carried out with respect to $x^\mu$ and $\sigma_s = -1$ if it were carried out with respect to $x'^\mu$ instead. To obtain $2 \delta^{(2)}[x-x']$ on the right hand side, the first term on the right hand side tells us we need to set $J[x=x']=1$ at the apex of the light cone $\sbar=0$. Moreover the second line on the right hand side must vanish. For the $\delta[\sbar]$ term of eq. \eqref{G_2DAnsatz_MainStep} to be zero, $(x-x')^\alpha \partial_\alpha J[\sbar=0] = 0$, i.e., $J$ is constant on the light cone and thus 
\begin{align}
\label{G_2D_LightConeBC}
J[\sbar=0]=1 
\end{align}
everywhere on the light cone. For the $\Theta[\sbar]$ term of eq. \eqref{G_2DAnsatz_MainStep} to be zero, the tail function $J$ needs to satisfy the homogeneous wave equation with respect to both $x^\mu$ and $x'^\mu$,
\begin{align}
\label{G_2D_WaveEqn}
\left(\partial^2 + U\right) J[\eta,\eta';\sbar] = 0 .
\end{align}
{\it Example} \qquad As an example, let us solve the flat spacetime symmetric Green's function $\mathcal{G}_2$ of the massive scalar, i.e.,
\begin{align}
(\partial^2 + m^2) \mathcal{G}_2[x-x'] = 2 \delta^{(2)}[x-x'] .
\end{align}
Because of the highly symmetric nature of the problem at hand, we will assume that $J$ depends on spacetime solely through $z \equiv m \sqrt{2\sbar}$. Then eq. \eqref{G_2D_WaveEqn} translates to
\begin{align}
(\partial^2 + m^2) J = m^2 \left(J''[z] + \frac{J'[z]}{z} + J[z] \right) = 0 ,
\end{align}
whose solutions are a linear combination of the Bessel functions $J_0[z]$ and $Y_0[z]$. To satisfy the light cone boundary condition in eq. \eqref{G_2D_LightConeBC}, we must discard $Y_0[z]$ because it is singular as $\sbar \to 0$,
\begin{align}
\mathcal{G}_2[x-x'] = \frac{\Theta[\sbar]}{2} J_0\left[ m \sqrt{2\sbar} \right] .
\end{align}
{\bf Tails and light cone B.C.'s} \qquad The tail of the Green's function is the term proportional to $\Theta[\sbar]$, describing the field propagating within the null cone, i.e., $\sbar > 0$. By applying the differential recursion relation in eq. \eqref{G_Recursion_Even} to the ansatz in eq. \eqref{G_2DAnsatz}, we see the Green's function tail in even dimensions is the term where all the $\sbar$-derivatives are acting on $J$, namely
\begin{align}
\label{G_Tail_Even}
\widehat{G}_{\text{even $d$}}^{\text{(tail)}}[\eta,\eta';\sbar] 
&= \frac{\Theta[\sbar]}{2 (2\pi)^{\frac{d-2}{2}}} 
\left(\frac{\partial}{\partial \sbar}\right)^{\frac{d-2}{2}} J[\eta,\eta';\sbar] .
\end{align}
We now obtain the light cone boundary condition for the tail function of the 4D and 6D Green's functions, namely $\partial_{\sbar} J[\sbar=0]$ and $\partial_{\sbar}^2 J[\sbar=0]$. The strategy is to exploit in 2D the light cone coordinates $x^\pm \equiv x^0 \pm x^1 = \eta \pm x^1$, so that 
\begin{align}
\dd s^2 = \dd x^+ \dd x^- , \qquad \partial_\pm = \frac{\partial_0 \pm \partial_1}{2} .
\end{align}
The homogeneous wave equation obeyed by $J$ in \eqref{G_2D_WaveEqn} becomes
\begin{align}
\label{G_2D_WaveEqnXPlusMinu}
\partial_+ \partial_- J = -\frac{1}{4} U J .
\end{align}
We employ a coordinate system such that $(x^+,x^-) = (x'^+,x'^-)$ defines the origin, and the light cone $\sbar = (1/2)(x^+ - x'^+)(x^- - x'^-) = 0$ consist of the positive-slope 45 degree line $x^- = x'^-$ and the negative-slope 45 degree line $x^+ = x'^+$. Because of the expected invariance of $J$ under parity, $\vec{x} \to -\vec{x}$ and $\vec{x}' \to -\vec{x}'$, in what follows we may focus on the $x^- = x'^-$ line (which is equivalent to $\eta-\eta'=x^1-x'^1$). Denote a prime as a derivative with respect to $\sbar$ and over dot one with respect to $\eta$. We will see how the $\partial_{\sbar} J[\sbar=0]$ and $\partial_{\sbar}^2 J[\sbar=0]$ can be solved once all the first and second derivatives with respect to the light cone coordinates are known. 

The first observation is that, since eq. \eqref{G_2D_LightConeBC} tells us $J[x^- = x'^-]=1$ for all $x^+$ and $J[x^+ = x'^+]=1$ for all $x^-$ we must have
\begin{align}
\label{G_2D_pd+All_LC}
\partial_+^{n \geq 1} J[x^- = x'^-] = 0, \qquad
\partial_-^{n \geq 1} J[x^+ = x'^+] = 0 .
\end{align}
This also means these derivatives are simultaneously zero at the origin, the apex of the light cone.
\begin{align}
\label{G_2D_pd+All_Apex}
\partial_+^{n \geq 1} J[x=x'] = \partial_-^{n \geq 1} J[x=x'] = 0 .
\end{align}
Central to our strategy is the wave equation \eqref{G_2D_WaveEqnXPlusMinu} evaluated on the light cone using eq. \eqref{G_2D_LightConeBC},
\begin{align}
\label{G_2D_WaveEqn_LC}
\partial_+ \partial_- J[x^- = x'^-] = -\frac{1}{4} U .
\end{align}
For, taking into account the apex boundary condition in eq. \eqref{G_2D_pd+All_Apex}, we may integrate with respect to $x^+$,
\begin{align}
\label{G_2D_pd-_LC_Integral}
\partial_- J[x^- = x'^-] 
= -\frac{1}{4} \int_{x'^+}^{x^+} \dd y^+ U\left[ ( y^+ + x^- )/2 \right] 
= -\frac{1}{2} \int_{\eta'}^{\eta} \dd\eta'' U\left[ \eta'' \right].
\end{align}
Acting $\partial_-$ on both sides of eq. \eqref{G_2D_WaveEqnXPlusMinu}, we have $\partial_+ \partial_-^2 J = -(1/4) ((1/2)\dot{U} + U \partial_- J)$; exploiting eq. \eqref{G_2D_pd-_LC_Integral}, we may again integrate with respect to $x^+$ to find
\begin{align}
\label{G_2D_pd--_LC_Integral}
\partial_-^2 J[x^- = x'^-]
&= \frac{U[\eta'] - U[\eta]}{4} + \frac{1}{4} \int_{\eta'}^{\eta} \dd \eta''  U\left[ \eta'' \right] \int_{\eta'}^{\eta''} \dd \eta''' U\left[\eta'''\right] .
\end{align}
On the other hand, via a direct calculation, we may also verify that the first derivatives are
\begin{align}
\label{G_2D_pd+_LC}
\partial_+ J[x^- = x'^-] &= \frac{1}{2} \dot{J} , \\
\label{G_2D_pd-_LC}
\partial_- J[x^- = x'^-] &= (\eta-\eta') J' + \frac{1}{2} \dot{J} .
\end{align}
The second derivatives are
{\allowdisplaybreaks\begin{align}
\label{G_2D_pd++_LC}
\partial_+^2 J[x^- = x'^-] 			&= \frac{1}{4} \ddot{J} , \\
\label{G_2D_pd--_LC}
\partial_-^2 J[x^- = x'^-] 			&= (\eta-\eta')^2 J'' + (\eta-\eta') \dot{J}' + \frac{1}{4} \ddot{J} , \\
\label{G_2D_pd+-_LC}
\partial_+ \partial_- J[x^- = x'^-] &= \frac{1}{2} (\eta-\eta') \dot{J}' + \frac{1}{2} J' + \frac{1}{4} \ddot{J} .
\end{align}}
Now, eq. \eqref{G_2D_pd+All_LC} applied to equations \eqref{G_2D_pd+_LC} and \eqref{G_2D_pd++_LC} indicates $\dot{J}[\sbar=0] = \ddot{J}[\sbar=0] = 0$ and therefore equations \eqref{G_2D_pd-_LC}, \eqref{G_2D_pd--_LC} and \eqref{G_2D_pd+-_LC} now yield
\begin{align}
J'[\sbar=0] 	&= \frac{\partial_- J[\sbar=0]}{\eta-\eta'} , \\
J''[\sbar=0] 	&= \frac{\partial_-^2 J[\sbar=0] + J'[\sbar=0] + U[\eta]/2}{(\eta-\eta')^2} .
\end{align}
Together with equations \eqref{G_2D_pd-_LC_Integral} and \eqref{G_2D_pd--_LC_Integral}, we may now gather
\begin{align}
\partial_{\sbar} J[\sbar=0] &= - \frac{1}{2(\eta-\eta')} \int_{\eta'}^{\eta} \dd \eta'' U\left[\eta''\right] , \\
\partial_{\sbar}^2 J[\sbar=0] &= \frac{1}{(\eta-\eta')^2} \Bigg\{ \frac{U[\eta] + U[\eta']}{4} - \frac{1}{2(\eta-\eta')} \int_{\eta'}^{\eta} \dd \eta'' U\left[\eta''\right] \\
&\qquad\qquad\qquad\qquad
+ \frac{1}{4} \int_{\eta'}^{\eta} \dd \eta'' U\left[ \eta'' \right] \int_{\eta'}^{\eta''} \dd \eta''' U\left[\eta'''\right] \Bigg\} . \nonumber
\end{align}
Although we will not pursue it further, it should be possible to continue this procedure to obtain the light cone boundary condition of the tail function in any even dimension, i.e., $\partial^{n \geq 1}_{\sbar} J[\sbar=0]$.

For the power law potential in eq. \eqref{PowerLawPotential}, the 4D and 6D tails obeys the light cone boundary conditions
\begin{align}
\label{G_4D_LCBC}
\text{4D : } \partial_{\sbar} J[\sbar=0] = \frac{p}{2 \eta\eta'} 
\end{align}
and
\begin{align}
\label{G_6D_LCBC}
\text{6D : } \partial_{\sbar}^2 J[\sbar=0] = \frac{p(p-2)}{8 (\eta\eta')^2} .
\end{align}
{\bf Power law potential} \qquad We now turn to solving the wave equation eq. \eqref{G_2D_WaveEqn} with the power law potential in eq. \eqref{PowerLawPotential}. To this end we postulate that $J$ depends on spacetime solely through the object $\sbar/(\eta\eta')$. (This is directly inspired by Nariai's ansatz in 4D \cite{Nariai}.) Doing so converts the wave equation \eqref{G_2D_WaveEqn} into an ordinary differential equation
\begin{align}
\left(\partial^2 - \frac{p}{\eta^2} \right) J[s]
= \frac{s (s+2) J''[s] + 2 (s+1) J'[s] - p J[s]}{\eta^2} = 0, \qquad s \equiv \frac{\sbar}{\eta\eta'} .
\end{align}
The general solution involves a linear combination of the Legendre functions $P_\nu[1+s]$ and $Q_\nu[1+s]$, where the $\nu$ depends on $p$. Because the light cone boundary condition in eq. \eqref{G_2D_LightConeBC} now corresponds to $J[s=0]=1$, and because $Q_\nu[1+s]$ blows up as $s \to 0$ while $P_\nu[1+s] \to 1$ in the same limit, the relevant solutions for the de Sitter analysis at hand are Legendre polynomials:
{\allowdisplaybreaks\begin{align}
\label{dS_J_Even_PseudoTrace}
J[s] &= P_{\frac{d-6}{2}} \left[1 + \frac{\sbar}{\eta\eta'}\right], \qquad\qquad p = \frac{(d-6)(d-4)}{4}, \\
\label{dS_J_Even_Vector}
J[s] &= P_{\frac{d-4}{2}} \left[1 + \frac{\sbar}{\eta\eta'}\right], \qquad\qquad p = \frac{(d-4)(d-2)}{4}, \\
\label{dS_J_Even_Tensor}
J[s] &= P_{\frac{d-2}{2}} \left[1 + \frac{\sbar}{\eta\eta'}\right], \qquad\qquad p = \frac{d(d-2)}{4}.
\end{align}}

\subsection{Odd Dimensions}

In three dimensions we first postulate the following ansatz for the symmetric Green's function:
\begin{align}
\label{G_3DAnsatz}
\widehat{G}_3[\eta,\eta';\sbar] \equiv \Theta[\sbar] J[x,x'] \equiv \Theta[\sbar] \frac{g_3[\eta,\eta';\sbar]}{2\pi \sqrt{2\sbar}} .
\end{align}
Because the symmetric minimally coupled massless scalar 3D Green's function obeys
\begin{align}
\partial^2 \left(\frac{\Theta[\sbar]}{2\pi \sqrt{2\sbar}}\right) = 2 \delta^{(3)}[x-x'] ,
\end{align}
and 
\begin{align}
	\partial^2 \left(\frac{1}{2\pi \sqrt{2\sbar}}\right) = 0  \qquad \text{ whenever }\qquad \sbar > 0 ;
\end{align}
plugging eq. \eqref{G_3DAnsatz} into the wave equation \eqref{G_MasterPDE} yields
\begin{align}
\label{G_3DAnsatz_MainStep}
(\partial^2 + U) \widehat{G}_3
&= 2 \delta^{(3)}[x-x'] g_3[x=x'] 
+ 2 \sigma_s \delta[\sbar]\left(\frac{(x-x')^\alpha \partial_\alpha g_3[\sbar=0]}{2\pi \sqrt{2\sbar}}\right)
+ \Theta[\sbar] (\partial^2 + U) J . 
\end{align}
where again $\sigma_s = +1$ if $\partial^2$ was carried out with respect to $x^\mu$ and $\sigma_s = -1$ if it were carried out with respect to $x'^\mu$ instead.  We require the right hand side to return $2 \delta[\eta-\eta'] \delta^{(2)}[\vec{x}-\vec{x}']$. For the coefficient of the $\delta$-function to be $2$, we need $g_3[x=x'] = 1$, on the apex of the light cone. For the $\delta[\sbar]$ term to vanish we have 
\begin{align}
\label{G_3D_LightConeBC_I}
\left(\frac{(x-x')^\alpha \partial_\alpha g_3}{\sqrt{\sbar}}\right)[\sbar=0] = 0 ,
\end{align}
i.e., $g_3$ is constant on the light cone $\sbar=0$. This in turn means $g_3$ is unity on the entire light cone:
\begin{align}
\label{G_3D_LightConeBC_II}
\lim_{\sbar \to 0} J[x,x']
= \lim_{\sbar \to 0} \frac{g_3[\sbar]}{2\pi \sqrt{2\sbar}} \to \frac{1}{2\pi \sqrt{2\sbar}} .
\end{align}
Moreover, demanding the vanishing of the $\Theta[\sbar]$ term in eq. \eqref{G_3DAnsatz_MainStep} informs us $J$ obeys the homogeneous wave equation:
\begin{align}
\label{G_3D_WaveEqn}
(\partial^2 + U)J = 0 .
\end{align}
{\it Example} \qquad As an example, let us solve the flat spacetime symmetric Green's function $\mathcal{G}_3$ of the massive scalar, i.e.,
\begin{align}
(\partial^2 + m^2) \mathcal{G}_3[x-x'] = 2 \delta^{(3)}[x-x'] .
\end{align}
Because of the highly symmetric nature of the problem at hand, we will assume that $J$ depends on spacetime solely through $z \equiv m \sqrt{2\sbar}$. Then eq. \eqref{G_3D_WaveEqn} translates to
\begin{align}
(\partial^2 + m^2) J = m^2 \left(J''[z] + \frac{2}{z} J'[z] + J[z] \right) = 0 ,
\end{align}
whose solutions are a linear combination of $\sin[z]/z$ and $\cos[z]/z$. To simultaneously require the vanishing of the light cone derivative of $g_3$ in eq. \eqref{G_3D_LightConeBC_I} and the light cone boundary condition in eq. \eqref{G_3D_LightConeBC_II}, we must ignore the $\sin[z]/z$ term and arrive at
\begin{align}
\mathcal{G}_3[x-x'] = \frac{\Theta[\sbar]}{2\pi \sqrt{2 \sbar}} \cos\left[ m \sqrt{2\sbar} \right] .
\end{align}
{\bf Tails} \qquad The tail of the Green's function is the term proportional to $\Theta[\sbar]$, describing the field propagating within the null cone, i.e., $\sbar > 0$. By applying the differential recursion relation in eq. \eqref{G_Recursion_Odd} to the ansatz in eq. \eqref{G_3DAnsatz}, we see the Green's function tail in odd dimensions is the term where all the $\sbar$-derivatives are acting on $J$, namely
\begin{align}
\label{G_Tail_Odd}
\widehat{G}_{\text{odd $d$}}^{\text{(tail)}}[\eta,\eta';\sbar] 
&= \frac{\Theta[\sbar]}{(2\pi)^{\frac{d-3}{2}}} 
\left(\frac{\partial}{\partial \sbar}\right)^{\frac{d-3}{2}} J[\eta,\eta';\sbar] \nonumber\\
&= \frac{\Theta[\sbar]}{(2\pi)^{\frac{d-3}{2}}} 
\left(\frac{\partial}{\partial \sbar}\right)^{\frac{d-3}{2}} \left(\frac{g_3[\eta,\eta';\sbar]}{2\pi \sqrt{2\sbar}}\right) .
\end{align}
{\bf Power law potential} \qquad We now derive explicit solutions to the wave equation \eqref{G_3D_WaveEqn} in the case where the potential is the power law in eq. \eqref{PowerLawPotential}. To this end we postulate the following Nariai-inspired ansatz \cite{Nariai}
\begin{align}
\label{G_J3_Ansatz}
J[x,x'] = \sqrt{H^2 a[\eta]a[\eta']} \widehat{J}[s], \qquad s \equiv \frac{\sbar}{\eta\eta'} .
\end{align}
When inserted into eq. \eqref{G_3D_WaveEqn}, this gives us the ordinary differential equation
\begin{align}
\left(\partial^2 - \frac{p}{\eta^2}\right) J 
= \frac{4 s (s+2) \widehat{J}''[s] + 12 (s+1) \widehat{J}'[s] + (3-4 p) \widehat{J}[s]}{4 \eta^2 \sqrt{\eta \eta'}} = 0 .
\end{align}
The solutions to $\widehat{J}$ are linear combinations of $P_{\nu-1/2}^{1/2}[1+s]/\sqrt[4]{s(s+2)}$ and $Q_{\nu-1/2}^{1/2}[1+s]/\sqrt[4]{s(s+2)}$, where the $P_{\nu-1/2}^{1/2}$ and $Q_{\nu-1/2}^{1/2}$ are associated Legendre functions, and $\nu$ depends on $p$. It turns out, to ensure the light cone derivative of $g_3$ goes to zero fast enough (eq. \eqref{G_3D_LightConeBC_I}), the $Q_{\nu-1/2}^{1/2}[1+s]/\sqrt[4]{s(s+2)}$ term needs to be discarded. For the de Sitter analysis at hand, where $p \in \{ (d-6)(d-4)/4, (d-4)(d-2)/4, d(d-2)/4 \}$, we invoke eq. 8.754.1 of \cite{G&S} to convert $P_{\nu-1/2}^{1/2}[\cosh \alpha] = (2/(\pi \sinh \alpha))^{1/2} \cosh[\nu\alpha]$ into an object built out of radicals of $s$. Specifically, upon matching onto \eqref{G_3D_LightConeBC_II}, one finds
\begin{align}
\label{dS_J_Odd_PseudoTrace}
J[x,x'] &= \frac{1}{4\pi} \frac{\left(s+\sqrt{s(s+2)}+1\right)^{d-5}+1}{\sqrt{\sbar (s+2)} \left(s+\sqrt{s (s+2)}+1\right)^{\frac{d-5}{2}}},
\qquad\qquad p \equiv \frac{(d-6)(d-4)}{4}, \\
\label{dS_J_Odd_Vector}
J[x,x'] &= \frac{1}{4\pi} \frac{\left(s+\sqrt{s(s+2)}+1\right)^{d-3}+1}{\sqrt{\sbar (s+2)} \left(s+\sqrt{s (s+2)}+1\right)^{\frac{d-3}{2}}},
\qquad\qquad p \equiv \frac{(d-4)(d-2)}{4}, \\
\label{dS_J_Odd_Tensor}
J[x,x'] &= \frac{1}{4\pi} \frac{\left(s+\sqrt{s(s+2)}+1\right)^{d-1}+1}{\sqrt{\sbar (s+2)} \left(s+\sqrt{s (s+2)}+1\right)^{\frac{d-1}{2}}},
\qquad\qquad p \equiv \frac{d(d-2)}{4} .
\end{align}
where $s$ is defined in eq. \eqref{G_J3_Ansatz}.

\subsection{Application to de Sitter}

With the symmetric Green's function $\widehat{G}$ of $(\partial^2 - p/\eta^2)$ at hand, the retarded solution to equations \eqref{dS_PseudoTrace_IIofII}, \eqref{dS_Vector_IIofII} and \eqref{dS_Tensor_IIofII} take the form (cf. eq. \eqref{G_MotivatingPDE})
\begin{align}
\label{G_MotivatingPDE_Solution}
\psi[\eta,\vec{x}] 
= -16\pi\GN \int_{-\infty}^{\eta} \dd\eta' \int_{\mathbb{R}^{d-1}} \dd^{d-1} \vec{x}' \left(\frac{a[\eta']}{a[\eta]}\right)^{\frac{d-2}{2}} \widehat{G}\left[\eta,\vec{x};\eta',\vec{x}'\right] S[\eta',\vec{x}'] .
\end{align}
For the de Sitter analysis, we have shown in the previous section that $\widehat{G}$ depends on $\sbar$ solely through the object $s \equiv \sbar/(\eta\eta')$; moreover, since $\eta\eta'>0$, we may assert $\Theta[\sbar] = \Theta[s]$. These allow us to apply the recursion eq. \eqref{G_Recursion_Even} for even $d \geq 4$ to eq. \eqref{G_2DAnsatz}, expressing all the $\sbar$-derivatives in terms of $s$-derivatives:
\begin{align}
\left( \frac{a[\eta']}{a[\eta]} \right)^{\frac{d-2}{2}} \left(\frac{\partial}{\partial \sbar}\right)^{\frac{d-2}{2}} \left( \frac{\Theta[\sbar]}{2} J[s] \right) &= H^{d-2} a[\eta']^{d-2} \left(\frac{\partial}{\partial s}\right)^{\frac{d-2}{2}} \left( \frac{\Theta[s]}{2} J[s] \right) .
\end{align}
Similarly, we may employ eq. \eqref{G_Recursion_Odd} for odd $d \geq 5$ to equations \eqref{G_3DAnsatz} and \eqref{G_J3_Ansatz},
\begin{align}
\left( \frac{a[\eta']}{a[\eta]} \right)^{\frac{d-2}{2}} \left(\frac{\partial}{\partial \sbar}\right)^{\frac{d-3}{2}} \left( \Theta[\sbar] \sqrt{H^2 a[\eta] a[\eta']} \widehat{J}[s] \right) 
&= H^{d-2} a[\eta']^{d-2} \left(\frac{\partial}{\partial s}\right)^{\frac{d-3}{2}} \left( \Theta[s] \widehat{J}[s] \right) .
\end{align}
We then surmise from equations \eqref{G_Recursion_Even} and \eqref{G_2DAnsatz}; \eqref{G_Recursion_Odd}, \eqref{G_3DAnsatz} and \eqref{G_J3_Ansatz}; that eq. \eqref{G_MotivatingPDE_Solution} becomes
\begin{align}
\label{Solution_FinalForm}
\psi[\eta,\vec{x}] 
&= -16\pi\GN H^{d-2} \int_{-\infty}^{\eta} \dd\eta' \int_{\mathbb{R}^{d-1}} \dd^{d-1} \vec{x}' a[\eta']^{d-2}
\mathcal{G}[s] S[\eta',\vec{x}'] ,
\end{align}
where
\begin{align}
\label{Solution_FinalForm_Even}
\mathcal{G}_{\text{even $d$}}[s] 
&\equiv \frac{1}{(2\pi)^{\frac{d-2}{2}}} \left(\frac{\partial}{\partial s}\right)^{\frac{d-2}{2}} \left( \frac{\Theta[s]}{2} J[s] \right) , \\
\label{Solution_FinalForm_Odd}
\mathcal{G}_{\text{odd $d$}}[s] 
&\equiv \frac{1}{(2\pi)^{\frac{d-3}{2}}} \left(\frac{\partial}{\partial s}\right)^{\frac{d-3}{2}} \left( \Theta[s] \widehat{J}[s] \right) .
\end{align}
For even $d \geq 4$, the explicit solutions of $J[s]$ are given in equations \eqref{dS_J_Even_PseudoTrace}, \eqref{dS_J_Even_Vector} and \eqref{dS_J_Even_Tensor}. For odd $d \geq 5$, the explicit solutions of $\widehat{J}[s]$ are given through equations \eqref{G_J3_Ansatz}, \eqref{dS_J_Odd_PseudoTrace}, \eqref{dS_J_Odd_Vector} and \eqref{dS_J_Odd_Tensor}. 

\section{Gauge invariant variables for perturbed spatially flat FLRW in various dimensions}
\label{Section_Bardeen}
The primary goal of this section is to identify the gauge-invariant metric variables of a perturbed $d$ dimensional spatially flat FLRW universe, generalizing Bardeen's in 4D. We first begin with the result that, under
\begin{align}
x^\alpha \to x^\alpha + \xi^\alpha ,
\end{align}
an arbitrary metric would transform as
\begin{align}
g_{\mu\nu} \to g_{\mu\nu} + \xi^\sigma \partial_\sigma g_{\mu\nu} + \partial_{ \{ \mu} \xi^\sigma g_{\nu \} \sigma} .
\end{align}
That means under $x \to x + \xi$, the perturbation of a conformally flat metric, namely the $\chi_{\mu\nu}$ in
\begin{align}
g_{\mu\nu}[\eta,\vec{x}] \equiv a[\eta]^2 \left( \eta_{\mu\nu} + \chi_{\mu\nu}[\eta,\vec{x}] \right) 
\end{align}
would transform as
\begin{align}
\label{GaugeTransformationOfChi}
\chi_{\mu\nu} 
\to \chi_{\mu\nu} + \partial_{\{\mu} \xi_{\nu\}} + 2 \frac{\dot{a}}{a} \xi^0 \eta_{\mu\nu} ,
\end{align}
where the indices on $\xi_\nu$ are moved with $\eta_{\alpha\beta}$. By exploiting the spatial rotational SO$_{d-1}$ symmetry of the background geometry, we shall perform an irreducible scalar-vector-tensor decomposition of the metric perturbation as well as an analogous one for $\xi$. That is, we do
\begin{align}
\xi_\mu = (\xi_0, \partial_i \ell + \ell_i), \qquad\qquad \delta^{ij} \partial_i \ell_j = 0 .
\end{align}
and
\begin{align}
\label{SVTDecomposition}
\chi_{00} \equiv E, \qquad \qquad \chi_{0i} \equiv \partial_i F + F_i, \nonumber\\
\chi_{ij} \equiv D_{ij} + \partial_{\{ i} D_{j\} } + \frac{D}{d-1} \delta_{ij} + \left( \partial_i \partial_j - \frac{\delta_{ij}}{d-1} \vec{\nabla}^2 \right) K ,
\end{align}
where these variables obey the following constraints:
\begin{align}
\label{Constraints}
\delta^{ij} \partial_i F_j = \delta^{ij} D_{ij} = \delta^{il} \partial_i D_{lj} = \delta^{ij} \partial_i D_j = 0 .
\end{align}
{\bf Gauge transformations} \qquad Applying the irreducible decomposition of eq. \eqref{SVTDecomposition} to eq. \eqref{GaugeTransformationOfChi}, under $x^0 \to x^0 + \xi_0$ and $x^i \to x^i - (\partial_i F + F_i)$, we may gather
\begin{align}
E \to E + 2 \frac{\partial_0 (a \xi_0)}{a} , \qquad\qquad
F \to F + \dot{\ell} + \xi_0, \qquad\qquad 
F_i \to F_i + \dot{\ell}_i,  \\
D_j \to D_j + \ell_j, \qquad\qquad
D \to D+2\vec{\nabla}^2 \ell - 2(d-1)\frac{\dot{a}}{a} \xi_0, \qquad\qquad
K \to K + 2\ell ,
\end{align}
and the transverse-traceless graviton is gauge-invariant
\begin{align}
D_{ij} \to D_{ij} .
\end{align}
{\bf Gauge invariant variables} \qquad At this point, we may then verify that the following variables are gauge invariant ones in a perturbed spatially flat $d$-dimensional FLRW universe. The $2$ scalar ones are
\begin{align}
\Psi &\equiv \frac{E}{2} - \frac{1}{a} \partial_0 \left\{ a \left( F - \frac{\dot{K}}{2} \right) \right\} , \\
\Phi &\equiv \frac{D - \vec{\nabla}^2 K}{d-1} + 2 \frac{\dot{a}}{a} \left( F - \frac{\dot{K}}{2} \right) ;
\end{align}
whereas the vector and tensor modes are, respectively,
\begin{align}
V_i \equiv F_i - \dot{D}_i
\qquad \qquad \text{ and }\qquad \qquad
D_{ij} \equiv \chi_{ij}^\text{TT} .
\end{align}
{\bf Remarks} \qquad In 2D the Einstein tensor is identically zero, and Einstein's eq. \eqref{Einstein} no longer defines dynamics for the metric. In 3D, the gauge-invariant tensor mode $D_{ij}$ does not exist. For, in momentum $\vec{k}$-space, if we choose a coordinate system such that $k^i = |\vec{k}| \delta^i_2$, the transverse condition in eq. \eqref{Constraints} would mean $0 = k^i D_{ij}[\vec{k}] = |\vec{k}| D_{j 2} = |\vec{k}| D_{2 j}$ and the traceless condition would imply $D_{11} = 0$; i.e., $D_{ij} = 0$. This is why, in this paper, we are restricting our attention to $d \geq 4$.

\end{document}